\documentclass[twocolumn]{aastex63}

\newcommand{\jwst}{JWST}
\newcommand{\hz}{high-$z$}
\newcommand{\lya}{Ly$\alpha$}
\newcommand{\ha}{H$\alpha$}
\newcommand{\hi}{H\,{\sc i}}
\newcommand{\buv}{$\beta$}
\newcommand{\muv}{$M_{\rm 1500}$}
\newcommand{\ewlya}{EW$_0$(Ly$\alpha$)}
\newcommand{\fescLya}{$f_{\rm esc}^{\rm Ly\alpha}$}
\newcommand{\fescLyC}{$f_{\rm esc}^{\rm LyC}$}
\newcommand{\ksi}{$\xi_{\rm ion}$}
\newcommand{\xio}{$\xi_{\rm ion, 0}$}
\newcommand{\Llya}{$L$(\lya)}

\newcommand{\Dd}{$\Delta d_{\rm Ly\alpha}$}

\newcommand{\wa}{FWHM(\lya)}

\newcommand{\hb}{H$\beta$}
\newcommand{\oiii}{[O\,{\sc iii}]}

\shorttitle{Luminous LAEs at $z\sim6$}
\shortauthors{Ning et al.}

\begin{document}
\title{An Updated Characterization of Luminous \lya\ emitters at the End of Reionization}

\author[0000-0001-9442-1217]{Yuanhang Ning}
\altaffiliation{yning@bjp.org.cn}
\affiliation{Department of Scientific Research, Beijing Planetarium, Beijing 100044, China}
\affiliation{Department of Astronomy, Tsinghua University, Beijing 100084, China}

\author[0000-0001-8467-6478]{Zheng Cai}
\altaffiliation{zcai@mail.tsinghua.edu.cn}
\affiliation{Department of Astronomy, Tsinghua University, Beijing 100084, China}

\author[0000-0003-4176-6486]{Linhua Jiang}
\affiliation{Kavli Institute for Astronomy and Astrophysics, Peking University, Beijing 100871, China}
\affiliation{Department of Astronomy, School of Physics, Peking University, Beijing 100871, China}

\author[0000-0002-3119-9003]{Qiong Li}
\affiliation{Jodrell Bank Centre for Astrophysics, University of Manchester, Oxford Road, Manchester M13 9PL, UK}

\author[0000-0002-3462-4175]{Si-Yue Yu}
\affiliation{Department of Astronomy, Xiamen University, Xiamen, Fujian 361005, China}

\author[0009-0004-1889-3043]{Xiaodi Yu}
\affiliation{Institute of Astronomy and Information, Dali University, Dali, Yunnan 671003, China}

\author[0000-0002-9634-2923]{Zhen-Ya Zheng}
\affiliation{CAS Key Laboratory for Research in Galaxies and Cosmology, Shanghai Astronomical Observatory, Shanghai 200030, China}

\begin{abstract}
We present a multi-wavelength physical characterization of 14 luminous \lya\ emitters (LAEs) at $z\approx6$, integrating deep ground-based Magellan/M2FS spectroscopy with heterogeneous JWST/NIRCam broad- and medium-band imaging. Identified via strong \lya\ lines with extreme \lya\ luminosities of ${>}10^{42.6}$~erg~s$^{-1}$, the sample exhibits very large rest-frame equivalent widths (${\gtrsim}100$~\AA) and steeply blue UV continua (\buv$_{\rm median}$ $\simeq-2.2$, $-18.2>$ \muv\ $>-20.2$ mag). 
Crucially, the integration of NIRCam medium-band photometry (F410M) breaks the degeneracy between strong rest-optical nebular emission and Balmer breaks, resolving probable mass overestimations. The tightly constrained spectral energy distribution modeling demonstrates that these luminous LAEs tend to be unequivocally low-mass, ultra-young dwarf starbursts; half the sample is characterized by stellar masses of $M_* < 10^8 M_{\odot}$, ages $\lesssim10$ Myr, and negligible dust attenuation. 
We also map the production efficiency of ionizing photons and Ly$\alpha$ escape fractions (\fescLya). 
The \fescLya\ values are exceptionally high, with a median of ${\gtrsim}40$\%, increasing for the bluer UV continua. Finally, analyzing spatial offsets between the \lya\ centroid and the stellar counterpart, we demonstrate empirically that internal dust content, rather than neutral hydrogen gas, dominate the suppression of \lya\ radiative transfer. Our study reveals that strong \lya\ emission of the luminous LAEs are generally attributed to both the vigorous starburst activities and the high \fescLya. Resembling Lyman continuum leakers, these extreme dwarf systems function as highly efficient ionizing engines at the conclusion of the Epoch of Reionization.

\end{abstract}


\section{Introduction}

\lya\ emitting galaxies are powerful probes to explore galaxy formation and evolution in the early universe.
Theoretically expected as a “spectral beacon” \citep{pp67}, \lya\ helps to gradually find galaxies at higher redshift \citep{hu96, rhoads00}. 
After the first galaxy to be spectroscopically confirmed by \lya\ at a redshift greater than 6 \citep{hu02}, \lya\ selected galaxies, the \lya\ emitters (LAEs) found by narrowband (NB) technique, keep spectroscopic redshift record as the most distant galaxies for a decade \citep[e.g.,][]{rhoads04, taniguchi05, iye06, martin08, tilvi10, hu10}. 

As a complementary method, the drop-out technique help select Lyman Break Galaxies (LBGs) or high-redshift (\hz) candidates \citep[e.g.,][]{steidel96a, giavalisco02, douglas09, bouwens15a, pentericci18b}. Especially, Hubble Space Telescope (HST) pushed the redshift frontier to $z\sim9$ with the installation of the Wide Field Camera 3 (WFC3/IR) \citep[e.g.,][]{ellis13, mcLeod16, oesch18, bouwens21}. After the launch in 2022, James Webb Space Telescope (JWST) substantially enlarges the sample at $z>9$ through deep infrared imaging \citep[e.g.,][]{castellano22, finkelstein22, finkelstein24, donnan23, franco24} and revolutionize the number of confirmed galaxies by the spectroscopic identification of \lya\ line and/or break \citep[e.g.,][]{arrabalharo23, bunker23, wangbj23, castellano24, carniani24}.

In the JWST era, galaxies at $z>7$ are being analyzed with \lya\ emission by spectroscopy in details \citep[e.g.,][]{chenzy24, napolitano24, jones24, tangm24b, heintz25, runnholm25, liq26}. The process of cosmic reionization becomes mostly complete at $z\lesssim6$ \citep[e.g.,][]{durovcikova24, spina24, zhuyd24}, which is a key stage to study how galaxies contribute to the ionizing budget. As potentially emitting Lyman Continuum (LyC) photons \citep[e.g.,][]{gazagnes20, choustikov24, izotov24, navarro-carrera25, 2026arXiv260202322M}, LAEs serve as an ideal laboratory at this redshift slice, especially those luminous ones which typically represent the extreme cases.

Although recent JWST works have specifically studied \hz\ LAEs \citep[e.g.,][]{goovaerts24, iani24}, most of the LAEs are not luminous and not representative of those already identified by the large ground-based surveys based on the NB technique \citep[e.g.,][]{kashikawa11, ning20, kikuta23}. 
On the other hand, some JWST works study the LAEs from ground-based surveys but most of them have not been spectroscopically confirmed, lacking the key redshift information \citep[e.g.,][]{shimizu25}.
More luminous (confirmed) LAEs thus need to be further analyzed, which could also in turn indirectly help preparing the candidates for the future ground-based surveys taken by, for example, the MUltiplexed Survey Telescope (MUST; \citealt{cai25, 2024arXiv241107970Z}).

We carried out a ground-based, multi-object Magellan/M2FS spectroscopic survey to establish a large sample of \hz\ galaxies \citep{jiang17}, including LAEs at $z\approx5.7$ \citep{ning20} and $z\approx6.6$ \citep{ning22}, and LBGs at $z\sim6$ \citep{fusq24}. 
These galaxies are spectroscopically confirmed via \lya\ lines from the \hz\ candidates which are designed to be selected from famous fields, waiting for JWST covering, including A370, CDFS, COSMOS, SXDS, and SSA22.
As JWST data is gradually accumulated by more surveys, more than 10 luminous LAEs from our M2FS survey have been covered by a series (${>}4$) of infrared (IR) bands, especially including the medium band (F410M). In this work, we make use of the imaging data to further explore these luminous LAEs, focusing on comparing their \lya\ properties and physical properties.

We organize this paper as follows. In Section 2, we briefly state the LAE sample from the M2FS spectroscopic observations, NIRCam imaging dataset from several JWST surveys, and photometric measurements. In Section 3, we describe the methods and present the results including the physical properties from spectral energy distribution (SED) fitting, UV and \ha-related quantities, \lya\ properties (line width, misalignment to stellar counterpart, escape fraction). We give further discussions in Section 4. We summarize our paper in Section 5. Throughout this work, we adopt a standard flat cosmology with $H_0=\rm{70\ km\ s^{-1}\ Mpc^{-1}}$, $\Omega_m=0.3$ and $\Omega_{\Lambda}=0.7$. All magnitudes refer to the AB system.

\begin{figure*}
\centering
\includegraphics[angle=0, width=1.\textwidth]{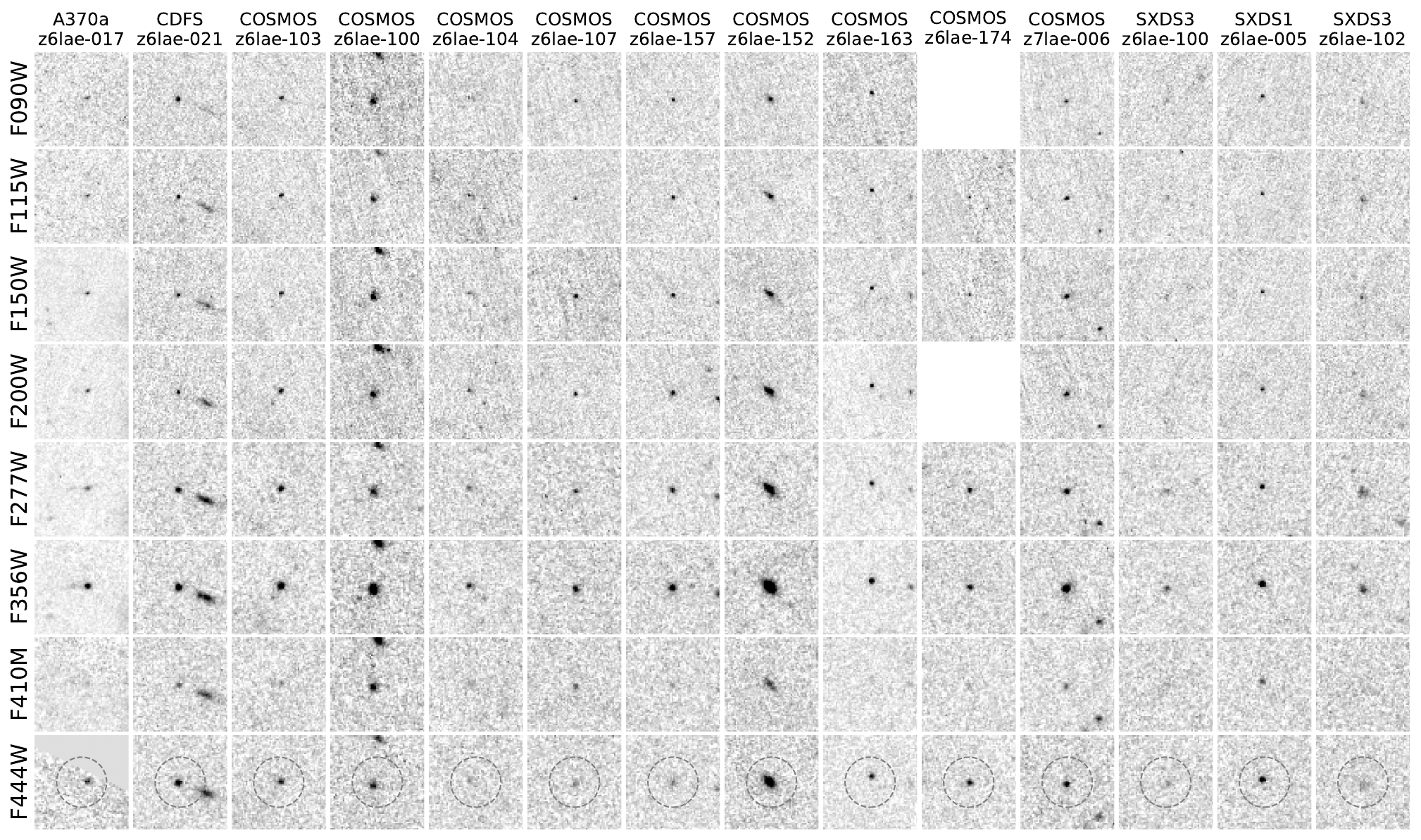}
\caption{Thumbnail images of the LAEs in our sample ($3\farcs0\times3\farcs0$, north is up and east to the left). Each LAE is shown with a top mark in each column. Each row shows images in different JWST/NIRCam bands indicated at the leftmost. In the lowest row (F444W), the dashed circles represent the $1\farcs6$ aperture centered at the \lya\ centroid. Note that COSMOS-z6lae-174 is not covered by F090W and F200W observations.
\label{samp}}
\end{figure*}

\section{Sample and Data}
In this section, we present the sample of the LAEs at $z\sim6$ with the JWST/NIRCam imaging data. Then we describe how to detect the sources and take photometry from the images for the LAE sample.

\subsection{Sample \& Imaging Data}

Our parent sample is drawn from the Magellan/M2FS spectroscopic survey of \hz\ galaxies presented in \citet{ning20, ning22}.
With the high-throughput spectrograph M2FS on the 6.5m Magellan Clay telescope, the effective integration time per pointing is averagely ${\sim}5$~hr (equivalently 35.8 hr in total for the sample in this work).
The LAEs are selected as candidates based on the Subaru/Suprime-Cam broadband and narrowband images and then identified with (strong) \lya\ lines by our Magellan/M2FS spectroscopic survey (in a \lya\ luminosity range from $10^{42.3}$ to $10^{43.7}$~erg s$^{-1}$). The survey spectroscopically confirmed 260 LAEs at $z\approx5.7$ and 36 LAEs at $z\approx6.6$ in total over nearly 2 deg$^2$ in the sky. 
So far as JWST has collected more extensive IR data across various fields, a fraction of our confirmed LAEs have been covered in more NIRCam bands. 
In this work, we retrieve and adopt the JWST imaging data from DAWN JWST Archive (DJA\footnote{\url{https://dawn-cph.github.io/dja/index.html}}), a repository of public JWST galaxy data from a number of JWST surveys. 
The JWST imaging data are reduced based on \texttt{grizli}\footnote{\url{https://zenodo.org/records/8370018}} \citep{brammer23}. \citet{valentino23} provides the NIRCam data reduction in detail. The version of the imaging dataset is as of December 2025.

Among the JWST/NIRCam bands, the medium ones are especially sensitive to the presence/absence of emission line. 
Including photometric data only covering optical continua can break the degeneracy between strong nebular lines of younger galaxies and prominent Balmer breaks of old galaxies \citep[e.g.,][]{jiang16a}. 
Thereinto, F410M is crucial to constrain continuum level for galaxies at $z\approx6$ while F410M$\--$F444W color is sensitive to \ha\ emission line of star-forming galaxies. 
To construct the sample, we thus select those covered by the NIRCam/F410M observations from the $260+36$ LAEs. Then we select those also detected in the F444W band and at least two bands among the F115W, F150W, and F200W bands. 
In this work, we focus on those detected only with a single counterpart (also see the end of next subsection). 
Among those satisfying the criteria, there is one in the A370 field and another one in the CDFS field, three in the SXDS field, and the rest nine locate in the COSMOS field. 

The final sample includes 13 LAEs at $z\approx5.7$ and a LAE at $z\approx6.6$.
All of them are luminous in terms of \lya, the lowest of which is ${\approx}10^{42.6}$~erg~s$^{-1}$. 
For comparison, two out of ${\lesssim}200$ LAEs in \citet{iani24} span a similar range of \lya\ luminosity as our sample (as shown by Figure \ref{sm}a), but their redshift values are all lower than 5.
In Figure~\ref{samp}, the thumbnail images in the JWST/NIRCam bands are shown for the LAE sample in this work. 
Most of them (12/14) are covered by the JWST Cycle-1 program (GO 1837), Public Release IMaging for Extragalactic Research (PRIMER; \citealt{primer}) 
which has the NIRCam imaging observations in eight bands, including F090W, F115W, F150W, and F200W in the short wavelength (SW), and F277W, F356W, F444W, and F410M in the long wavelength (LW). They are also covered by the \jwst\ Cycle 1 program (GO 1727), COSMOS-Web \citep{cosmosWeb, casey23}.
CDFS-z6lae-021 is covered by the JWST Cycle-1 program (GO 1210), JWST Advanced Deep Extragalactic Survey (JADES/South; \citealt{jades23, jades26}). A370a-z6lae-017 is covered by the JWST Cycle-1 program (GO 1208), The CAnadian NIRISS Unbiased Cluster Survey (CANUCS; \citealt{willott22, canucsTDR1}). 
Note that COSMOS-z6lae-101 is excluded because its JWST colors and morphology suggest that it is likely an “Little Red Dot” (LRD) source, as discussed in our previous work \citet{ning24}.
A370a-z6lae-024 is also excluded because it is not as luminous as the sample in terms of \lya\ luminosity, which is magnified ($\mu\gtrsim10$) by the foreground cluster A370 \citep{hc17}.

\subsection{Photometry}

We use \texttt{SEP} \citep{barbary16}, a python implementation of \texttt{SExtractor} \citep{ber96}, to detect sources and measure the photometry of the objects in the bands. 
We create the empirical point-spread function (PSF) by selecting and stacking at least 100 bright (not saturated) stars for each band. The images are PSF-homogenized to F444W and stacked to feed the detection image.
We then take the source detection by adopting a minimum area of five pixels, a threshold of 2σ, 64 deblending levels, and a contrast of 0.001 with a gaussian kernel and match the output catalogs to the original \lya\ centroid (from the NB imaging data). In the candidate selection of our Magellan/M2FS program, the photometric apertures have an angular radius of $1\farcs0$. The spectroscopic fibers have an angular radius of $0\farcs6$. We set the distance tolerance by a radius of $0\farcs8$, averaging the above two values. 
We do not perform forced photometry at the NB-detected positions due to the possible misalignment between \lya\ emission and stellar counterparts as report by pervious works \citep[e.g.,][]{ning24}.

Next we also use \texttt{SEP} to perform photometry in all bands for each detected source. This procedure is same as the dual-image mode in \texttt{SExtractor}. 
For the signal-to-noise ratio (S/N) of faint object detection while maintaining the accuracy of total flux, we measure the total \texttt{MAG\_AUTO} flux with elliptical apertures as did by previous works \citep[e.g.,][]{jiang20a}. 
Kron factor and the minimum aperture size are set to be 2.0 and 2.5, respectively. 
The aperture corrections are also measured by increasing both the parameters,
which is equivalent to the standard \texttt{PHOT\_AUTOPARAMS} values of 2.5 and 3.5. 
Our sample only includes the sources with a single counterpart detected above $3\sigma$ in the F444W band. Such single counterpart dominates \ha\ emission and contributes \lya\ emission so that it is proper to obtain the corresponding \lya\ escape fraction when comparing other properties.

\section{Results}

In this section, we first give the results of the physical properties based on the SED fitting. We then constrain the UV continua using the photometric data, estimate \ha\ lines with the best-fit models from the SED fitting and calculate their production efficiency of ionizing photons. Next we measure their \lya\ properties, including the line width, offset to stellar counterpart, and escape fraction. We list the results in Table \ref{tab}. We compare the potential relationship between them in Figure~\ref{comp}.

\begin{figure*}[t]
\centering
\gridline{\fig{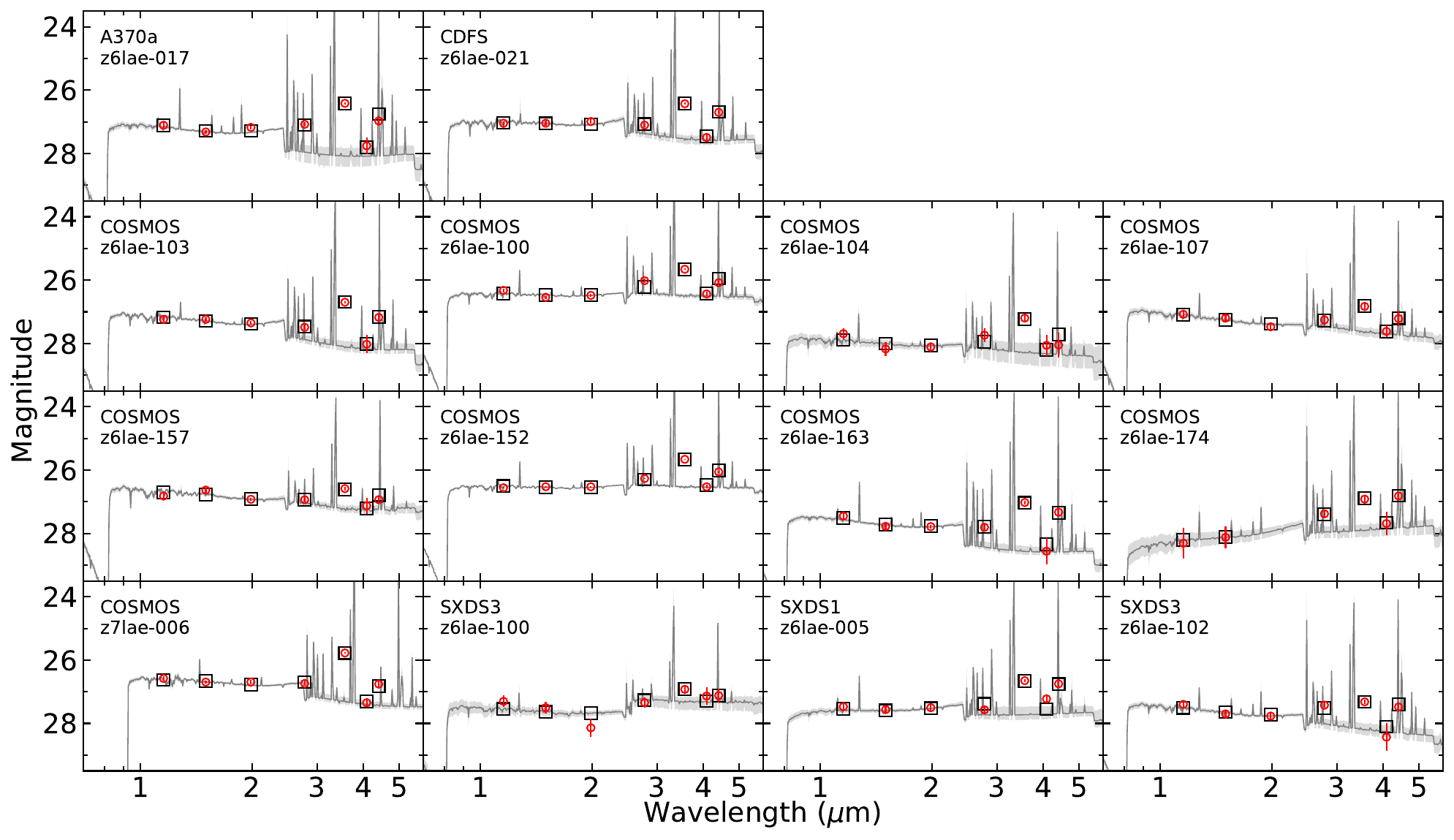}{0.95\textwidth}{(a)}}
\gridline{\fig{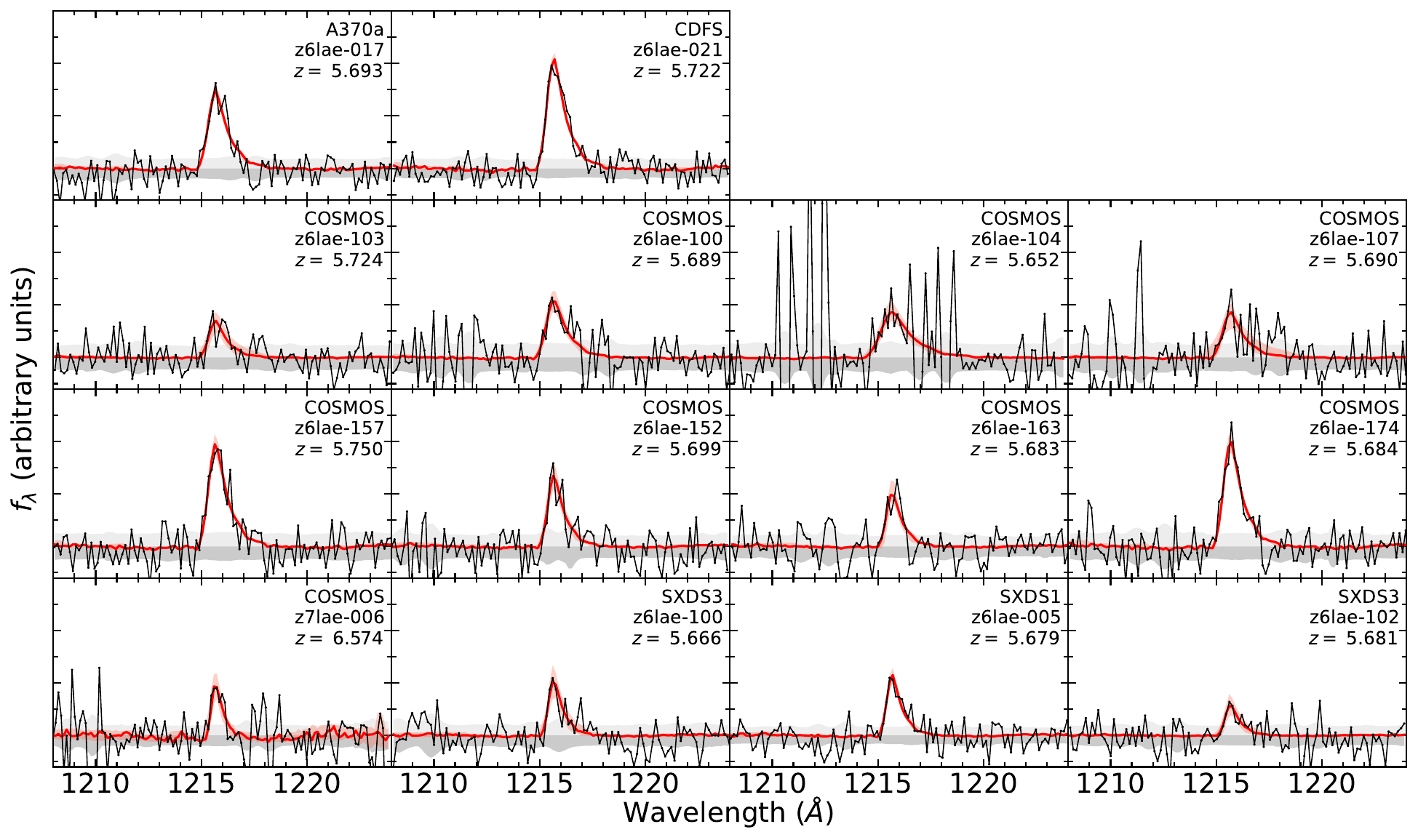}{0.95\textwidth}{(b)}}
\caption{\textbf{(a)} SED fitting in the (rest-frame) UV to optical bands for the LAEs in our sample. In each panel for each LAE, the gray line represents the best-fit SED model with the light gray region representing $1\sigma$ uncertainty. The red circles and black squares are the photometric data points from JWST/NIRCam imaging observations and computations of the best-fit mode, respectively.
\textbf{(b)} \lya\ lines in the Magellan/M2FS spectra of LAEs in our sample. Flux density $f_{\lambda}$ is shown in arbitrary units. The (slightly different) gray-shaded regions indicate $(\pm) 1\sigma$ uncertainty level. The red lines are the best-fit \lya\ profiles based on the line templates while the light shaded region represent the 16–84th percentiles ($1\sigma$) of the posterior.
\label{fit}
}
\end{figure*}

\subsection{Physical Properties}

To constrain and obtain the physical properties, we perform SED fitting using \texttt{BAGPIPES}, a Python framework
for self-consistently modeling spectra and photometry with galactic properties \citep{carnall18}. We run the latest version with the pure Python \texttt{Nautilus} nested sampling algorithm \citep{lange23}. In the fitting, we do not include the F090W data due to the \lya\ line strongly boosting this band. We do not also consider the \lya\ line in the posterior models of spectra. Flux of this line is mainly up to magnitude of the corresponding NB data as stated as follows.

For our sample, their strong \lya\ emission indicates the ongoing star formation activity dominating the observed SEDs. Considering the limited number of photometric datapoints, we employ a constant star-forming history (SFH) model like previous studies on \hz\ LAEs \citep[e.g.,][]{jiang20a, simmonds23, begley24}. We start to perform the SED fitting within a broad parameter space. The age varies logarithmically from 0.1 Myr to the cosmic age at the redshift of each source (${\sim}1$~Gyr). The formed mass is set in the logarithmic range of $6-11$. The metallicity also varies logarithmically within the range of 0.001 and 10 times the solar metallicity. The nebular emission has a Cloudy model with an ionization parameter of $-4<{\rm log}(U)<0$. 
Assuming a Calzetti law \citep{calzetti00} for dust attenuation, $A_V$ (absolute attenuation in the V band) is assigned to vary between 0 and 4 magnitudes. The stellar/nebular attenuation ratio is set to be unity as suggested by previous studies on \hz\ star-forming galaxies \citep[e.g.,][]{faisst19}.

The SED fitting results are shown with $1\sigma$ uncertainty region in Figure~\ref{fit}(a). Derived physical properties (Age, $A_V$, and $M_*$) are listed in the Column 5, 6, and 7 of Table \ref{tab} (the upper part), respectively. 
LAEs generally tend to be young galaxies \citep[e.g.,][]{jiang16a, iani24, firestone25}, which is confirmed by the output properties of the LAE sample in this work. 
Strikingly, half of them are young with ages smaller than 10 Myr. 
Only one has an age exceeding 100 Myr.
For most of the LAEs in the sample, their small $A_V$ show the low content of dust ($A_V<1$) while only one has $A_V>0.5$. The median value of $A_V$ is 0.12. The output $M_*$ has a logarithmic range of ${\sim}7.5-8.5$ and a median of ${<}8.0$. 
We give a further discussion about the results of stellar mass in Section 4.1.

\floattable
\renewcommand{\arraystretch}{1}
\begin{splitdeluxetable*}{rccccccBrcccccc}
\tabletypesize{\normalsize}
\tablecaption{Measured properties of the LAEs in the sample of this work.
\label{tab}}
\centering
\tablehead{
   \colhead{ID} & 
   \colhead{$\beta$} & 
   \colhead{$M_{\rm 1500}$} & 
   \colhead{lg $ L$(\ha)} & 
   \colhead{Age} &
   \colhead{$A_V$} & 
   \colhead{lg $M_*$} & 
   \colhead{ID} & 
   \colhead{lg $\xi_{\rm ion,0}$} &
   \colhead{lg $L$(\lya)} & 
   \colhead{lg \ewlya} & 
   \colhead{\Dd} & 
   \colhead{\fescLya} &
   \colhead{\wa}
   \\
   \colhead{} & 
   \colhead{} &
   \colhead{} &
   \colhead{(erg s$^{-1}$)} & 
   \colhead{(Gyr)} & 
   \colhead{} &
   \colhead{($M_{\odot}$)} & 
   \colhead{} & 
   \colhead{(Hz erg$^{-1}$)} &
   \colhead{($\rm erg\ s^{-1}$)} & 
   \colhead{(\AA)} & 
   \colhead{(kpc)} & 
   \colhead{} &
   \colhead{(km s$^{-1}$)} \\
   \colhead{(1)} & \colhead{(2)} & \colhead{(3)} & \colhead{(4)} & \colhead{(5)} & \colhead{(6)} & \colhead{(7)} & \colhead{(1)} & \colhead{(8)} & \colhead{(9)} & \colhead{(10)} & \colhead{(11)} & \colhead{(12)} & \colhead{(13)}
   }
\startdata
A370a-z6lae-017 & -1.90 $\pm$ 0.43 & -19.50 $\pm$ 0.16 & 42.39 $\pm$ 0.07 & $0.005_{-0.003}^{+0.004}$ & $0.076_{-0.076}^{+0.130}$ & $7.62_{-0.08}^{+0.15}$ & A370a-z6lae-017 & 25.75 $\pm$ 0.09 & 42.92 $\pm$ 0.03 & 2.27 $\pm$ 0.09 & 0.48 $\pm$ 0.29 & 0.39 $\pm$ 0.06 & $223.58_{-23.05}^{+17.29}$ \\
CDFS-z6lae-021 & -1.91 $\pm$ 0.05 & -19.70 $\pm$ 0.12 & 42.42 $\pm$ 0.06 & $0.006_{-0.002}^{+0.003}$ & $0.240_{-0.074}^{+0.062}$ & $7.72_{-0.08}^{+0.10}$ & CDFS-z6lae-021 & 25.52 $\pm$ 0.07 & 43.00 $\pm$ 0.02 & 2.28 $\pm$ 0.07 & 0.06 $\pm$ 0.26 & 0.44 $\pm$ 0.06 & $220.75_{-12.31}^{+14.10}$ \\
COSMOS-z6lae-103 & -2.21 $\pm$ 0.10 & -19.52 $\pm$ 0.12 & 42.22 $\pm$ 0.06 & $0.006_{-0.002}^{+0.003}$ & $0.087_{-0.046}^{+0.059}$ & $7.48_{-0.05}^{+0.10}$ & COSMOS-z6lae-103 & 25.55 $\pm$ 0.07 & 42.62 $\pm$ 0.03 & 1.95 $\pm$ 0.06 & 0.47 $\pm$ 0.47 & 0.29 $\pm$ 0.04 & $243.71_{-58.82}^{+96.14}$ \\
COSMOS-z6lae-100 & -2.20 $\pm$ 0.31 & -20.20 $\pm$ 0.09 & 42.60 $\pm$ 0.05 & $0.023_{-0.009}^{+0.014}$ & $0.348_{-0.051}^{+0.048}$ & $8.39_{-0.15}^{+0.14}$ & COSMOS-z6lae-100 & 25.39 $\pm$ 0.07 & 42.96 $\pm$ 0.01 & 2.03 $\pm$ 0.05 & 0.81 $\pm$ 0.27 & 0.26 $\pm$ 0.03 & $233.62_{-29.16}^{+35.70}$ \\
COSMOS-z6lae-104 & -2.67 $\pm$ 0.44 & -18.72 $\pm$ 0.31 & 41.91 $\pm$ 0.14 & $0.012_{-0.008}^{+0.065}$ & $0.158_{-0.134}^{+0.181}$ & $7.54_{-0.31}^{+0.53}$ & COSMOS-z6lae-104 & 25.49 $\pm$ 0.18 & 42.71 $\pm$ 0.04 & 2.36 $\pm$ 0.17 & 1.23 $\pm$ 0.64 & 0.72 $\pm$ 0.23 & ... \\
COSMOS-z6lae-107 & -2.70 $\pm$ 0.16 & -19.53 $\pm$ 0.17 & 41.97 $\pm$ 0.14 & $0.026_{-0.016}^{+0.031}$ & $0.004_{-0.004}^{+0.091}$ & $7.88_{-0.25}^{+0.24}$ & COSMOS-z6lae-107 & 25.38 $\pm$ 0.15 & 42.70 $\pm$ 0.02 & 2.00 $\pm$ 0.10 & 0.41 $\pm$ 0.40 & 0.62 $\pm$ 0.20 & ... \\
COSMOS-z6lae-157 & -2.29 $\pm$ 0.43 & -19.98 $\pm$ 0.13 & 42.12 $\pm$ 0.10 & $0.026_{-0.014}^{+0.023}$ & $0.046_{-0.046}^{+0.100}$ & $8.05_{-0.20}^{+0.19}$ & COSMOS-z6lae-157 & 25.30 $\pm$ 0.11 & 42.63 $\pm$ 0.04 & 1.75 $\pm$ 0.09 & 0.35 $\pm$ 0.54 & 0.38 $\pm$ 0.09 & $223.82_{-14.26}^{+19.35}$ \\
COSMOS-z6lae-152 & -1.95 $\pm$ 0.03 & -20.12 $\pm$ 0.04 & 42.58 $\pm$ 0.02 & $0.022_{-0.005}^{+0.005}$ & $0.378_{-0.023}^{+0.026}$ & $8.38_{-0.09}^{+0.07}$ & COSMOS-z6lae-152 & 25.37 $\pm$ 0.03 & 42.98 $\pm$ 0.01 & 2.09 $\pm$ 0.03 & 0.74 $\pm$ 0.26 & 0.29 $\pm$ 0.02 & $192.54_{-25.24}^{+37.26}$ \\
COSMOS-z6lae-163 & -2.51 $\pm$ 0.25 & -19.09 $\pm$ 0.14 & 42.16 $\pm$ 0.06 & $0.003_{-0.001}^{+0.002}$ & $0.020_{-0.020}^{+0.077}$ & $7.34_{-0.04}^{+0.06}$ & COSMOS-z6lae-163 & 25.73 $\pm$ 0.08 & 42.70 $\pm$ 0.03 & 2.21 $\pm$ 0.07 & 1.27 $\pm$ 0.41 & 0.40 $\pm$ 0.06 & $163.55_{-25.81}^{+30.69}$ \\
COSMOS-z6lae-174 & -1.35 $\pm$ 0.24 & -18.30 $\pm$ 0.54 & 42.51 $\pm$ 0.10 & $0.040_{-0.029}^{+0.180}$ & $0.758_{-0.192}^{+0.179}$ & $8.47_{-0.43}^{+0.41}$ & COSMOS-z6lae-174 & 25.65 $\pm$ 0.24 & 42.87 $\pm$ 0.02 & 2.79 $\pm$ 0.28 & 0.04 $\pm$ 0.33 & 0.26 $\pm$ 0.06 & $213.66_{-18.12}^{+19.00}$ \\
COSMOS-z7lae-006 & -2.18 $\pm$ 0.14 & -20.29 $\pm$ 0.10 & 42.58 $\pm$ 0.10 & $0.003_{-0.002}^{+0.003}$ & $0.157_{-0.059}^{+0.069}$ & $7.85_{-0.06}^{+0.06}$ & COSMOS-z7lae-006 & 25.54 $\pm$ 0.10 & 43.09 $\pm$ 0.06 & 2.12 $\pm$ 0.08 & 0.32 $\pm$ 0.28 & 0.37 $\pm$ 0.10 & ... \\
SXDS3-z6lae-100 & -3.30 $\pm$ 0.46 & -19.03 $\pm$ 0.36 & 41.82 $\pm$ 0.17 & $0.242_{-0.199}^{+0.383}$ & $0.021_{-0.021}^{+0.219}$ & $8.57_{-0.43}^{+0.25}$ & SXDS3-z6lae-100 & 25.42 $\pm$ 0.22 & 42.81 $\pm$ 0.05 & 2.34 $\pm$ 0.19 & 0.53 $\pm$ 0.42 & 1.13 $\pm$ 0.45 & $183.21_{-45.55}^{+55.42}$ \\
SXDS1-z6lae-005 & -2.03 $\pm$ 0.15 & -19.03 $\pm$ 0.19 & 42.49 $\pm$ 0.08 & $0.003_{-0.002}^{+0.005}$ & $0.432_{-0.098}^{+0.110}$ & $7.73_{-0.06}^{+0.14}$ & SXDS1-z6lae-005 & 25.66 $\pm$ 0.11 & 42.84 $\pm$ 0.04 & 2.42 $\pm$ 0.10 & 0.17 $\pm$ 0.35 & 0.26 $\pm$ 0.05 & $178.35_{-16.61}^{+22.05}$ \\
SXDS3-z6lae-102 & -2.62 $\pm$ 0.22 & -19.19 $\pm$ 0.16 & 41.99 $\pm$ 0.09 & $0.035_{-0.024}^{+0.091}$ & $0.007_{-0.007}^{+0.108}$ & $7.86_{-0.38}^{+0.40}$ & SXDS3-z6lae-102 & 25.53 $\pm$ 0.11 & 42.73 $\pm$ 0.05 & 2.17 $\pm$ 0.10 & 0.47 $\pm$ 0.40 & 0.64 $\pm$ 0.15 & $190.36_{-52.90}^{+82.81}$ \\
\enddata
\centering
\tablecomments{“lg” denotes the base-10 logarithm.}
\end{splitdeluxetable*}

\subsection{UV Continua and \ha\ Lines}

We use the output SED-fitting model spectra to estimate UV continua of the sample, from which we have the absolute UV magnitude \muv\ at the rest-frame wavelength of 1500 \AA.
We constrain the UV-continuum slope with the three SW photometric datapoints (F115W, F150W, and F200W; only F115W and F150W for COSMOS-z6lae-174). We do not make use of the F090W data which is largely contributed by \lya\ emission. We keep using a power-law formula as did by previous works \citep[e.g.,][]{jiang20a, ning23, ning24}. We fit the datapoints by a linear relation (in AB magnitude units) of $m_{\rm AB} \propto (\beta+2)\times{\rm log}(\lambda)$ to obtain the UV slope \buv.

We also estimate the flux of \ha\ line based on the output SED-fitting model spectra. For each LAE, we interpolate the continuum around \ha\ line and subtract it to obtain the \ha\ flux. 
The corresponding flux error comes from the $1\sigma$ uncertainty range of the \ha-flux ensemble from posterior SED models obtained by fitting the photometric datapoints with the photometric uncertainties.
Note that for the galaxies at $z\sim6$, without constraining the optical continua, the model spectra usually degenerate between strong nebular lines (of young galaxies) and prominent Balmer breaks (of old galaxies). The F410M photometric data breaks such degeneracy and significantly improves the accuracy of estimating the \ha\ line flux.

After constraining UV continua and \ha\ line, we compute the Hydrogen ionizing photon production efficiency \ksi\ by
\begin{eqnarray}
   \xi_{\rm ion} = \frac{\dot N_{\rm ion}}{L_{\nu}^{\rm UV}}.
\end{eqnarray}
In this formula, $L_{\nu}^{\rm UV}$ (erg~s$^{-1}$~Hz$^{-1}$) is the mono-chromatic UV-continuum luminosity per photon frequency. It is usually derived from the measured \muv. $\dot N_{\rm ion}$ (s$^{-1}$) is the intrinsic production rate of Hydrogen ionizing photons from stellar populations, which can be calculated from \ha\ emission by 
\begin{eqnarray}
\dot N_{\rm ion} = \frac{L({\rm H\alpha})}{1-f_{\rm esc}^{\rm LyC}} \times 7.35\times10^{11}\ {\rm erg^{-1}}
\end{eqnarray} in the ($T_e=10^4 K$) case-B recombination \citep{kennicutt94, lh95, madau98}. 
We then correct $L_{\nu}^{\rm UV}$ and $L({\rm H\alpha})$ with the stellar/nebular extinction ratio of unity and the dust reddening law of \citet{calzetti00}.
We adopt the production efficiency of ionizing photons \xio\ by assuming \fescLyC\ = 0. 

\begin{figure*}[htp]
\centering
\includegraphics[angle=0, width=1\textwidth]{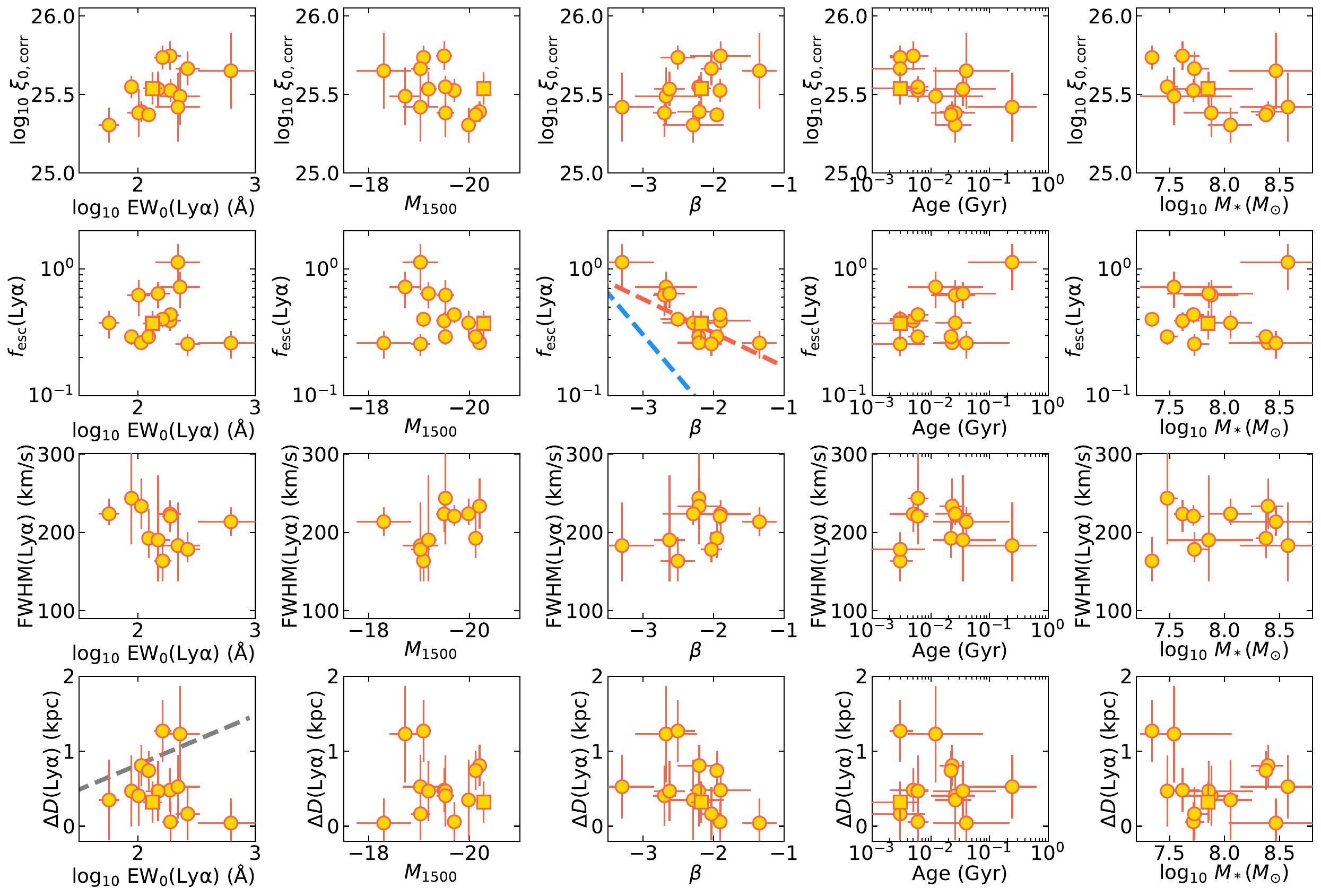}
\caption{Comparison between a series of properties of the LAE sample in this work. 
The square represents the LAE at $z\approx6.6$ in our sample.
In the \fescLya-\buv\ panel, the red and blue dashed lines indicate best-fit linear relations based on the LAE sample in this work and a (\lya-emitting) LBG sample from our previous work \citep{ning23}, respectively.
In the \Dd-\ewlya\ panel, the dashed line gives the best-fit linear relation from our previous work \citep{ning24}.
\label{comp}}
\end{figure*}

In Table~\ref{tab}, Column 8 lists the \xio\ results.
The derived UV and \ha\ properties are plotted in Figure~\ref{comp}. The results show that the LAEs have a median \muv\ of ${\simeq}-19.5$ mag in the range of [--20.2, --18.2] mag, which are slightly fainter than those of \citet{ning24}. This is because the surveys covering the sample in this work are deeper than the large-area COSMOS-Web survey used in our previous work.
The UV slopes of our sample have a range roughly from --3.3 to --1.3. The median value is \buv\ $\simeq-2.2$, which is basically consistent with those from the literature \citep[e.g.,][]{jiang13a, 2026arXiv260119995J, 2026arXiv260120045A}. Comparing with the brighter or fainter sample \citep[e.g.,][]{jiang13a, 2026arXiv260120045A} suggests no clear dependence between UV slopes and luminosities for the LAEs at the redshift.

The \xio\ results distribute in a range of log$_{10}$\,\xio~$\sim25.2-25.8$, which is supposed to be a typical \xio\ range of luminous LAEs at $z\approx6$. 
The \xio\ range is consistent with that of a LBG sample (representing \hz\ star-forming galaxies) from our previous work \citet{ning23} if we exclude SC-1 therein which has a probability as a LRD of $\sim$ 80\% with a F277W–F444W color of ${\sim}1.5$ \citep{greene24}. 
In Figure~\ref{comp}, a positive correlation exists between \xio\ and \ewlya\ which is also reported by \citet{ning23}. 
However, there is no clear (anti-)correlation between \xio\ and \buv\ like the results reported by literature works \citep[e.g.,][]{castellano23, saldanaLopez23}.
Such difference implies the possibility of different interstellar medium (ISM) physical conditions of the low-mass galaxies associated to the LAEs. 
The plausibly positive \xio-\buv\ trend is similar to that from a recent work \citet{2026arXiv260120045A} which demonstrate the extra contribution by nebular continuum emission with no significant variation in the stellar population of very faint star-forming galaxies. 
Despite a source with \buv\ reaching $-1$, our sample has slightly wider \buv\ range. Although nebular continuum may contribute to the observed trend, the range with $\Delta\beta>1$ is still less expected from nebular continuum alone so additional factors related to nebular conditions, such as \fescLyC, are probably required.

\subsection{\lya\ Properties}

\subsubsection{\lya\ Line Width}

For the LAE sample in this work, we have the \lya\ spectra taken from the Magellan/M2FS Spectroscopic Observations. 
More details are provided in \citet{jiang17} and \citet{ning20}.
To measure the width of \lya\ lines, we implement a Markov-Chain-Monte-Carlo method in the line fitting. 
We adopt the dimensionless line profile of our composite \lya\ template from our previous works \citep{ning20, ning22}.
For each LAE, we have the spectroscopic redshift. We vary scaling factors of line flux from 0.3 to 3.0 times the peak value. We shrink and expand the template line to adjust its width with scaling factors in a range of [0.2, 5].

In Figure~\ref{fit}(b), we plot the fitting results for all sources in the sample. 
We can see that fitting is suitable for most of the sources but bad for COSMOS-z6lae-104 and COSMOS-z6lae-107. The reason is that there are too many “spikes” around \lya\ wavelength. If we mask these abnormal values, the number of data points is not enough for executable fitting especially at the redder side of \lya\ peak. We thus abandon these two sources when comparing \lya\ line width with other properties. 
For COSMOS-z7lae-006, its \lya\ profile is considerably cut by the skyline subtraction at the red side of the peak so we ignore its line width result. 
For the \lya\ lines in this work, we adopt the observed FWHMs without the correction of the instrument broadening. On the one hand, the instrumental resolution is different for each spectrum but can be treated to be approximately constant (varies slightly around $R\sim2000\pm50$) in the narrow NB816 wavelength range. On the other hand, in this work, we aim to only compare their relative values with each other within our sample, rather than with those from other studies.
In Table~\ref{tab}, Column 13 lists the results of \wa. We plot the \wa\ as functions of other properties in the 3rd row of Figure~\ref{comp}. We do not found any clear relation between \wa\ and these properties for our LAE sample.

\subsubsection{Ly$\alpha$ Offset from Stellar Counterpart}

Recent JWST observations show the non-negligible misalignment between \lya\ (or LyC) emission and stellar counterparts \citep[e.g.,][]{jianghc24, napolitano24, ning24, 2026arXiv260305907W}. 
As shown by Figure~\ref{samp}, the stellar counterparts of the LAEs revealed by JWST/NIRCam do not locate at the corresponding \lya\ centroid, the center of the dashed circular aperture. We have reported such phenomenon of the misalignment between \lya\ emission and the stellar counterpart of luminous LAEs in our previous work \citet{ning24}. In this work, we explore how such misalignment relates to several properties. The \lya\ and stellar centroids are detected still based on the Subaru/Suprime NB images and the JWST/NIRCam images, respectively. The positional distances are computed to obtain the \lya-stellar projected offsets \Dd.
We follow the same method and procedure as \citet[see the Section 3.4]{ning24} to measure and evaluate the corresponding errors.

The astrometric accuracy is first addressed. We use the detected sources within a circle of 2\farcm0 diameter and compare their positions in the JWST/NIRCam and  Suprime NB images. The systematic offsets are obtained by computing the median and eliminated from \Dd. The astrometric precision is also addressed. We perform a Monte Carlo simulation by inserting the mock LAEs randomly in the NB images. We obtain the positional error of source detections as a function of LAE brightness. Fainter LAEs have larger measurement errors. For each source in the sample, the error of \Dd\ is generated by interpolating its NB photometric data to the above magnitude-dependent relations. More details are provided in the Section 3.4 of \citet{ning24}.
 
In Table~\ref{tab}, Column 11 lists the results of \Dd. We plot the \Dd\ as functions of other properties in the bottom row of Figure~\ref{comp}. Part of them spread consistently along the best-fit linear function previously obtained by \citet[see Figure 5]{ning24} but others locate below this relation. The projection effect is supposed to be a main cause. If \lya\ emission goes along the line of sight when escaping from a galaxy, such observed offset is still very small even though the true offset is large. 
We also discuss how \Dd\ is affected by other properties in next section.

\subsubsection{\lya\ Escape fraction}

Before computing \fescLya, we update the \lya\ luminosities of our spectroscopically confirmed LAEs in this work. The flux of \lya\ line is re-computed based on the NB photometric data and the best-fit SED model for each LAE. Except \lya\, the continuum around Lyman break also contributes the NB magnitude although the \lya\ line usually dominates. We subtract the continuum integral flux (weighted by the NB filter curve) from the total NB flux. 
Then by matching the above difference, we obtain the \lya\ flux with the above mentioned template of the composite \lya\ line which is redshifted to the observed frame for each one. We then estimate the \lya\ equivalent width in the rest frame, \ewlya.
Note that the spectroscopic redshift is crucial when measuring \lya\ line flux with NB data. NB filter curve is largely different from the top-hat shape. A strong line may exhibits faint if it is (red)shifted into the edge of the NB filter curve.

With the above obtained flux of \ha\ line, we compute the escape fraction of \lya\ photons (\fescLya) for each LAE. We adopt the canonical ratio $L({\rm Ly\alpha})/L({\rm H\alpha})=8.7$ under the assumption of case-B recombination in $T_e=10^4 K$ \citep{agn2} to calculate the intrinsic \lya\ flux and obtain: 
\begin{eqnarray}
f_{\rm esc}^{\rm Ly\alpha} = \frac{L_{\rm obs}({\rm Ly\alpha})}{L({\rm H\alpha})\times8.7}
\end{eqnarray} where $L_{\rm obs}({\rm Ly\alpha})$ is the observed \lya\ luminosity and $L({\rm H\alpha})$ is the intrinsic \ha\ luminosity corrected by the dust reddening law of \citet{calzetti00}. Although lacking information on the Balmer decrement (\ha/\hb), we assume that \texttt{BAGPIPES} provides proper $A_V$ values in the SED-fitting output. In Table~\ref{tab}, Column 12 lists the results of \fescLya. We plot the computed \fescLya\ as functions of other properties in the 2rd row of Figure~\ref{comp}. 

For our LAE sample, \fescLya\ is all larger than ${\sim}20\%$ (a stacked result of \citealt{2026arXiv260306409T}) which could be a lower limit of the luminous LAEs at $z\approx6$. 
For these luminous LAEs, we do not see any indication of the relation between \ewlya\ and \fescLya. Literature works report the existence of the positive relation for the \lya-emitting galaxies at low- or \hz\ \citep[e.g.,][]{yang17b, kimk21, ning23, 2025arXiv250918302P}. In these previous works, most of galaxies in the sample have \ewlya\ $<100$ \AA. For the sample in this work, most of the luminous LAEs have \ewlya~$>100$~\AA\ 
in a limited dynamic range in \ewlya. Note that half of the sample have \fescLya~$\gtrsim40\%$, which is broadly consistent with the result of \citet{sobral19a}.

On the whole, our obtained \fescLya\ results are larger than the star-forming galaxies at a similar redshift range from other works \citep[e.g.,][]{ning23, chenzy24, tangm24a}.
Although no apparent relation is found between \fescLya\ and \ewlya, such difference suggests the luminous LAEs from the NB technique, with relatively high \ewlya, tend to own larger \fescLya\ than \lya-emitting galaxies primarily selected by other methods such as the drop-out technique. 
Another reason could be possible. Part of literature works carry out JWST/NIRSpec prism observations for \lya\ \citep[e.g.,][]{chenzy24, heintz25}, which account for \lya\ flux lose due to limited size of the slits (and the placement) and the non-negligible \lya\ offset to the stellar counterpart \citep{jianghc24, ning24}.
Our results of \fescLya\ are obtained based on the photometric data, which is less influenced by the above problems.

\begin{figure*}[t]
\centering
\gridline{\fig{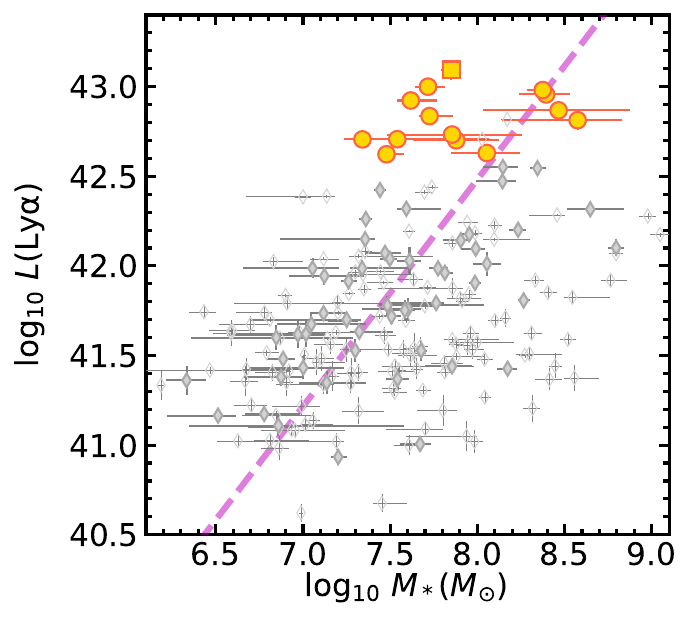}{0.4\textwidth}{(a)}
\fig{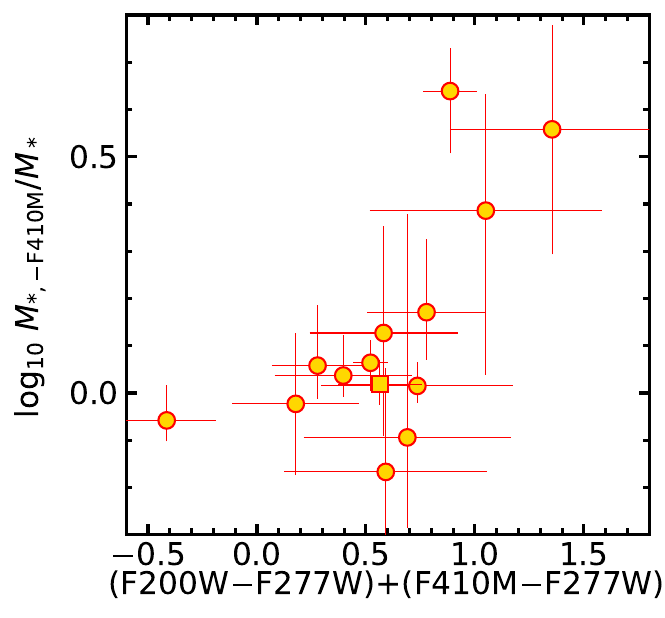}{0.385\textwidth}{(b)}}
\caption{\textbf{(a)} 
\Llya\ as a function of $M_*$: besides our sample indicated by the same markers, the filled and open gray diamonds represent the less luminous sample of LAEs at $z>4.5$ and $z\leq4.5$ from \citet{iani24}, respectively. The dashed line is the best-fit linear function to the $z>4.5$ subsample.
\textbf{(b)} 
Comparing the stellar masses derived from the SED fitting with the inclusion ($M_*$) and exclusion ($M_{*, \rm -F410M}$) of F410M: the horizontal axis represents the sum of two colors, F200W$-$F277W and F410M$-$F277W.
\label{sm}
}
\end{figure*}

We do not find a clear relation between \fescLya\ and \muv. 
For a negative trend of \fescLya\ changing as \buv, we obtain a linear relation 
\begin{equation}
{\rm log_{10}}\,f_{\rm esc}^{\rm Ly\alpha} = (-0.27 \pm 0.08)\,\beta_{\rm UV} - (1.04 \pm 0.17)
\label{eq:fescbuv}
\end{equation}
as plotted by a red dashed line in Figure~\ref{comp}. This relation is higher than that obtained by our previous work (plotted by a blue dashed line in Figure~\ref{comp}) overall by ${\sim}0.5$ dex. The two trends diverge as \buv\ goes redder, which illustrates \fescLya\ is less affected for the luminous LAEs than other (drop-out) galaxies emitting \lya\ despite the \fescLya\ difference demonstrated by the preceding paragraph.
Among these LAEs, one LAE SXDS3-z6lae-100 has a large \fescLya\ reaching ${\sim}100\%$ with a very steep UV slope of \buv~$\approx-3$. 
But note that the errors of \fescLya\ and \buv\ are both relatively large because it is faint and its \ha\ line is also weak. 
This LAE is similar to a very blue starburst galaxy with \fescLyC~$\sim1$ at $z\sim10$ reported by \citet{2026arXiv260202322M}. 
\fescLyC\ is usually smaller than \fescLya\ \citep[e.g.,][]{dijkstra16, izotov20}. 
The ${\gtrsim}20\%$ \fescLya\ of luminous LAEs optimistically imply their potential modest \fescLyC\ and contribution to the budget of ionizing photons to cosmic reionization.

\begin{figure*}
\centering
\includegraphics[angle=0, width=0.86\textwidth]{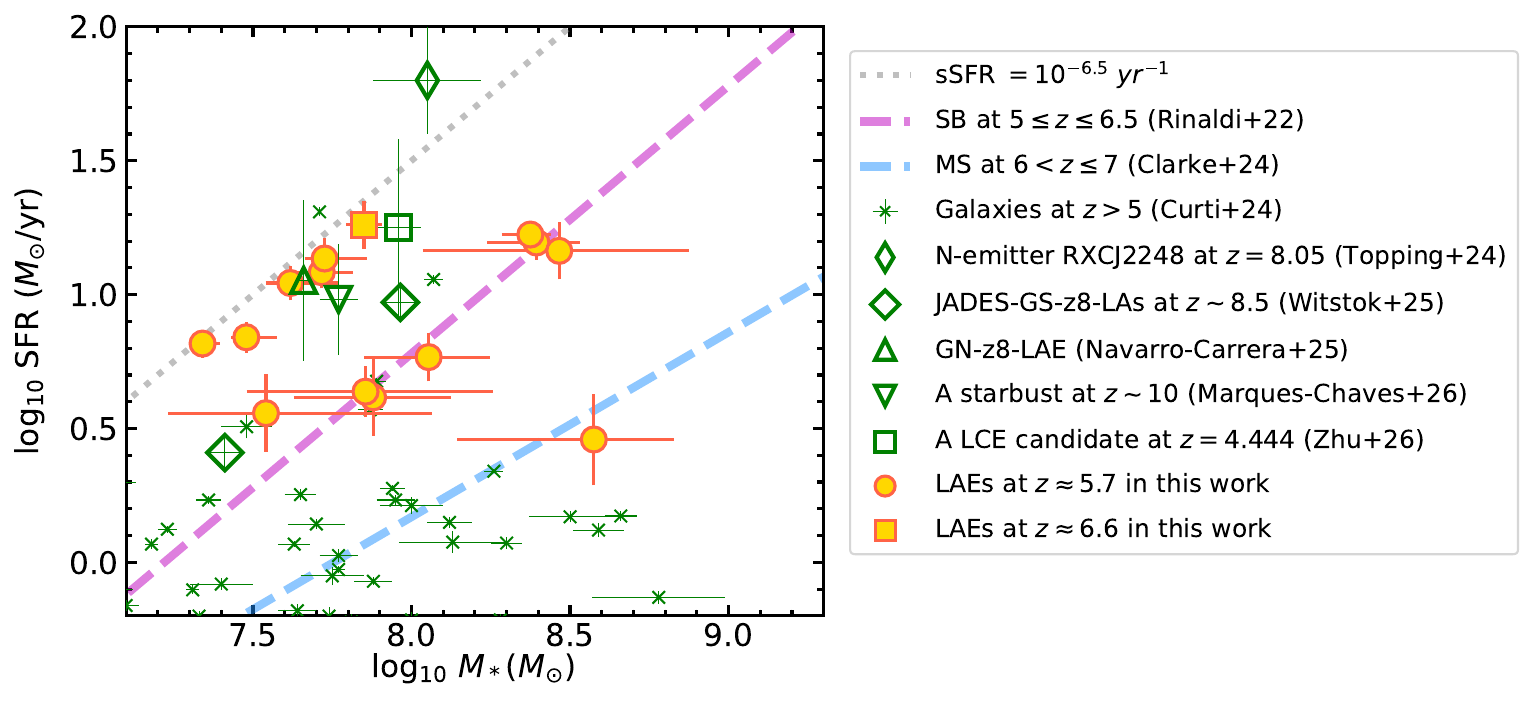}
\caption{SFR vs. $M_*$ diagram:
The blue dashed line show the best-fit SFMS at $6 < z \leq 7$ obtained by \citet{clarke24}.
The magenta dashed lines represent the best-fit power-law relations from \citet{rinaldi22}. 
The crosses indicate the $z>5$ galaxies from \citet{curti24}. 
The following open markers also represent the results from the literature.
The square is LCEz4-M1, a Lyman continuum emitter (LCE) candidate at $z = 4.444$ reported by \citet{2026arXiv260301487Z}.
The thin diamond is RXCJ2248, a N-emitter at $z = 8.05$ from \citet{topping24}.
The thick diamonds are JADES-GS-z8-0-LA and JADES-GS-z8-1-LA, respectively, two $z\sim8.5$ LAEs with \fescLya\ $\sim10\%$ from \citet{witstok25}.
The upward triangle is GN-z8-LAE, an LAE $z \approx 8.3$ from \citet{navarro-carrera25}. 
The downward triangle is a very blue starburst galaxy with \fescLyC~$\sim1$ at $z\sim10$ reported by \citet{2026arXiv260202322M}. 
\label{sfr}}
\end{figure*}

\section{Discussion}
In this section, we discuss the results of stellar mass and its possible relation to the \lya\ luminosity. We next discuss the bursty activity of star formation for our LAE sample. 
We also explore the relation between \Dd\ and other derived properties.

\subsection{About Stellar Mass}
The LAEs in our sample have the derived stellar masses in a logarithmic range of ${\sim}7.5-8.5$. The median value is lower than $10^{8.0} M_{\odot}$.
Previous works show the luminous LAEs, identified by the ground-based observations, have a typical stellar mass range of log$_{10} M_*/M_{\odot} \gtrsim 8-9$ \citep[e.g.,][]{jiang16a, ouchi20, ning24}.
With the help of more near-IR bands, a series of the JWST/NIRCam bands in this work, our derived $M_*$ of the luminous LAEs spread over a reduced range by a factor of ${\sim}0.5$ dex.
In Figure \ref{sm}(a), we plot the \lya\ luminosity as a function of stellar mass to compare our sample with a less luminous LAE sample from \citet[][hereafter I24]{iani24}. 
Their sample has a broad redshift range over $z=2.8-6.7$. 
We split it to mainly use a higher-redshift subsample at $z>4.5$ (filled gray diamonds) to compare our sample at $z\sim6$. 
The \lya\ luminosities of our sample are all brighter than $10^{42.5}$~erg s$^{-1}$ while the I24 subsample of LAEs at $z>4.5$ has a broad \Llya\ range. 
The I24 subsample exhibits a broad \Llya-$M_*$ trend, which is natural consider the largely scattering \fescLya\ and the SFR (namely \ha)-$M_*$ relation for star-forming galaxies.
Our LAEs basically follow and locate slightly higher which indicates larger \fescLya\ and/or more bursty star formation intrinsically. 

As we mentioned earlier, the inclusion of medium bands (in the rest optical wavelength range) helps better estimating the stellar mass by underlining the rest-optical continuum or capturing the nebular emission lines \citep[e.g.,][]{robertsBorsani21, ning23, lyuc25}. 
Without the medium band(s), performing SED fitting only with broad bands (mixing stellar continuum and nebular emission lines) likely gives a larger Balmer break with weaker nebular lines instead of originally strong emission lines, and vice versa. 
For our sample at $z\sim6$, the medium band F410M, covering the optical continuum between \oiii+\hb\ and \ha, plays a crucial role to constraining the stellar mass. 
In Figure \ref{sm}(b), we compare the derived stellar masses from the SED fitting with the inclusion and exclusion of F410M.
The result apparently exhibits an overall mass difference, demonstrating the probable mass overestimation after excluding F410M.
Especially three of them have difference factors reaching ${\sim}0.5$ dex. Part of them have little difference owing to the relatively abundant broad bands.
Also note that each LAE is confirmed by the strong \lya\ line for our sample thus removing one critical free parameter "redshift". 
Their spectroscopic redshifts assure that nebular emission lines locate in proper bands.

We further investigate the probable overestimation of the derived stellar mass without F410M by comparing the photometric information.
For our sample, we empirically find a plausible correlation between the $M_*$ difference and the sum of two colors, F200W$-$F277W and F410M$-$F277W. For the source COSMOS-z6lae-174, we adopt F150W for the lacking F200W. When the two colors have a larger sum, the $M_*$ difference would be larger. The reason can be intuitively explained as follows. The color F200W$-$F277W can represent Balmer break while F410M$-$F277W corresponds to the optical continuum color. The stronger Balmer break and bluer optical continuum indicate larger and smaller mass-to-light ratios, and then larger difference of the inferred stellar masses (at fixed rest-optical luminosity), respectively \citep[e.g.,][]{zibetti09, vikaeus24}. When the sum of two colors goes larger, the derived $M_*$ probably increase without the information of optical continuum if excluding F410M.

\subsection{Bursty Star Formation}

We compare the star-formation rates (SFRs) with stellar mass ($M_*$) for our sample of luminous LAEs at $z\sim6$.
These LAEs with \ewlya\ $\gtrsim50$ \AA\ (12/14 $>100$ \AA) are associated to very low-mass galaxies. We use the method of \citet{kennicutt98b} adapted for \citet{chabrier03} IMF to compute the SFR based on the \ha\ luminosity. 
Most of the LAEs are very young with the SED-derived ages of $\lesssim30$ Myr. 
“Recent” star-formation activities power nebula to emit (optical) lines, \ha\ emission can infer SFRs averaged over a timescale of 10 Myr revealed by recent works using state-of-the-art hydrodynamical simulations \citep[e.g.,][]{2025arXiv250905403K, 2026arXiv260105916I}. We compare the inferred SFRs based on the \ha\ luminosities with those derived from SED fitting. They have overall consistency within around 0.1 dex. We plot the \ha-based SFRs as a function of the stellar mass by comparing the results from literature in Figure~\ref{sfr}. 

In the SFR-$M_*$ diagram, the luminous LAEs tend to own starburst activity. 
Most of galaxies at $z>5$ \citep{curti24} follow the SFMS relation \citep{clarke24} while our sample has a relatively diverse distribution. 
Only one LAE in our sample locates closed to the MS line and half of them lie around the relation derived for starburst galaxies for $z\sim5-6.5$ by \citet{rinaldi22}. 
It is worth noting that our LAEs have (overall) lower mass than \hz\ extreme emission line galaxies from \citet{2025arXiv251025647L}.
With the seven NIRCam (6 broad- and 1 medium-) bands in the SED fitting, we constrain their physical properties more accurately compared to our previous work \citet{ning24}. One parameter of the crucial properties, stellar mass, is overall reduced with a median of ${\rm log}_{10} M^* < 8.0$, which exhibit the importance to utilize the medium band covering the (rest-frame) optical continuum, as discussed by Section 4.1.

Strikingly, another half of the LAEs exhibit apparently more vigorous bursts of star formation, beyond the starburst regime. 
These LAEs even have higher sSFR reaching ${\approx}10^{6.5} \rm{yr}^{-1}$, larger than the two UV-bright LAEs at $z\sim8.5$ \citep{witstok25} and 
the starburst branch corresponding to the young LAEs at $z\simeq3-7$ reported by \citet{iani24}.
As shown by Figure~\ref{sfr}, they also resemble LCEz4-M1, GN-z8-LAE and U37126 in the aspects of age, SFR, and $M_*$. 
LCEz4-M1 is a LyC emitter candidate at $z = 4.444$ which is determined by the \lya\ line \citep{2026arXiv260301487Z}.
GN-z8-LAE is a prominent LAE at $z\approx8.3$ and a robust LyC leaker candidate with a strong C\,{\sc iv}] line \citep{navarro-carrera25}.
U37126 is an strong LyC leaker at $z\approx10.3$ but without \lya\ line due to a phase of ISM-naked starburst \citep{2026arXiv260202322M}. 
Considering strong \lya\ lines as one of the reliable indicators of LyC emission \citep[e.g.,][]{gazagnes20, izotov24}, such similarity may suggest our luminous LAEs efficiently leak LyC emission supporting ionized bubbles in the Epoch of Reionization (EoR), which is also our conclusion from our previous work \citet{ning22} by comparing \lya\ luminosity functions at redshift 5.7 and 6.6.

\subsection{\lya\ Transfer and Escape}

Our LAE sample exhibits non-negligible misalignment between \lya\ emission and stellar counterparts. We also reveal such phenomenon in our previous work \citet{ning24}. We discussed a tentative relation between \Dd\ and \ewlya. For luminous LAEs, \lya\ emission would displace farther with larger \ewlya. In this work, we constrain physical properties of the LAEs better with photometric data from more JWST/NIRCam bands. We can explore whether other properties determine \Dd\ more effectively.

Unlike other emission lines such as \ha, \lya\ suffer resonant scattering by neutral Hydrogen (\hi) atoms and passes longer routines before absorbed by dust \citep[e.g.,][]{maoj07, dijkstra11, dijkstra14}. \lya\ emission thus distributes more complex in/around galaxies \citep[e.g.,][]{hayes15}. Besides gas turbulence or outflow altering the pathways of \lya\ emission \citep[e.g.,][]{riveraThorsen15}, possible low-opacity channels in the ISM/CGM, driven by Supernova (SN) feedback \citep[e.g.,][]{2026arXiv260211261K} and/or AGN \citep[e.g.,][]{wangwj24}, may also help \lya\ and ionizing photons leaking \citep[e.g.,][]{2025arXiv251213778M}. \lya\ photons thus propagate inhomogeneously and anisotropically so that its overall distribution could misalign with the corresponding stellar population.

\begin{figure}
\epsscale{1.13}
\centering
\plotone{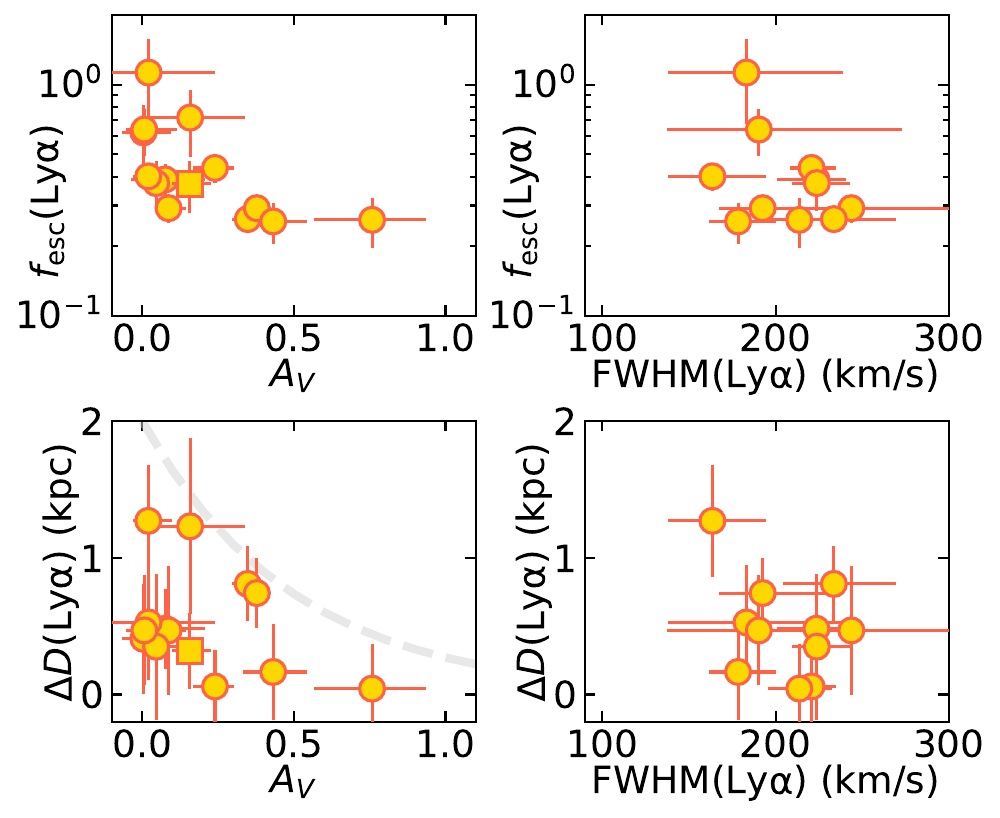}
\caption{\lya\ escape fraction (\fescLya) and offset to stellar counterparts (\Dd) as functions of dust attenuation ($A_V$) and \lya\ line width \wa. In the lower left panel, we manually plot a dashed line indicating a limit trend of suppressing \Dd\ as $A_V$ increasing.
\label{lyatran}}
\end{figure}

In Figure~\ref{lyatran}, we compare \fescLya\ and \Dd\ with $A_V$ and \wa, respectively. The plots shows that both \fescLya\ and \Dd\ are obviously suppressed by increasing $A_V$ but much less influenced when changing \wa. The trends can be explained by the mechanism of \lya\ transfer process. $A_V$ quantitatively measures the attenuation of dust content. The \lya\ line width due to Doppler or (multiple) scattering broadening, expressed by \wa\ in this work, is directly influenced by the density, temperature, and kinematic state of the \hi\ atoms \citep{dijkstra14}. 

In general, more \hi\ would increase times of resonant scattering and distances of propagation for \lya\ photons, which boosts velocity offset with respect to the systemic redshift $\Delta\nu_{\rm Ly\alpha}$ and broadens the \lya\ line width \citep[e.g.,][]{dijkstra16, 2025arXiv250918302P}. However, \hi\ gas content can not determine how much \lya\ photons survive out of galaxies and how far they reach when scattered to velocities where the gas is transparent. 
Although literature works report broad \hi\ (absorption) velocity width results in lower \fescLya, this anti-correlation is demonstrated by a higher probability for \lya\ photons to be destroyed by dust grains when more likely residing in the optically thick \hi\ gas \citep[e.g.,][]{gazagnes20}.

If a galaxy contains more dust content, \lya\ emission would be less possible to escape and propagate far away due to being absorbed by dust grains \citep{dijkstra14}. This trend is qualitatively represented by the dashed line in the lower left panel. The dust content thus plays a relatively dominant role in regulating \lya\ transfer and escape. 
Its \Dd\ could also be small if \lya\ is emitted homogeneously and/or the (spatial) centroid locates closed to the line of sight.
The above demonstration can be verified further with the $\Delta\nu_{\rm Ly\alpha}$ data measured from the difference of observed wavelength between \lya\ line and any other optical lines such as \oiii, \ha\ and so on.

\section{Summary}

In this work, we study a spectroscopically confirmed sample of 14 luminous LAEs in terms of \lya\ luminosity ($>10^{42.6}$ erg s$^{-1}$) over a large sky area. 
In this sample, each LAE is selected if associating with a single counterpart of the star-forming galaxy. 
We utilize the imaging and spectroscopic datasets including Subaru/Suprime NB optical images, Magellan/M2FS multi-object fiber spectra, and JWST/NIRCam IR images in six broad bands and one medium band. 
By performing SED fitting, we constrain their physical properties (stellar mass, age, and dust attenuation) and estimate flux of \ha\ lines. We also obtain the UV and \ha\ properties (UV slope and ionizing photon production efficiency), and \lya\ properties (line width, \lya\ offset, and escape fraction). This work is summarized as follows:

\begin{itemize}
 \item{The luminous LAEs in the sample have very high \ewlya. Most of them have \ewlya~$>~100$\AA. They span an absolute UV magnitude (\muv) range from --18.2 to --20.2 mag. They are generally blue in an UV slope range of \buv~$\sim[-3.3, -1.3]$ with a median of \buv\ $\simeq-2.2$.
 }
 \item{For our sample, these luminous LAEs are generally very young, in which half of them have ages roughly smaller than 10 Myr. They have low dust attenuate with a median $A_V\approx0.12$. Their stellar mass is low in a range of $7.3<{\rm log}_{10} M_*<8.6$ with an median of ${\rm log}_{10} M_* < 8.0$, which basically follows but locates slightly higher than a broad \Llya-$M_*$ trend of less luminous LAEs.
 }
 \item{Our sample has a \xio\ range of log$_{10}$\,\xio~$\sim25.2-25.8$ basically coinciding with star-forming galaxies at the same redshift range. Except the understandable positive correlations of \xio-\ewlya\ and \xio-\muv, we do not find a clear negative relation between \xio\ and \buv\ as reported by previous works.
 }
 \item{The luminous LAEs in the sample all have \lya\ escape fractions of \fescLya~$>0.2$. Half of them have \fescLya~$\gtrsim0.4$. For the sample, a negative correlation still exist between \buv\ and \fescLya\ (higher than other star-forming galaxies): those with bluer UV slopes tend to have larger \fescLya\ (Equation (\ref{eq:fescbuv})).
 }
 \item{The inclusion of NIRCam medium-band photometry (F410M in this work) could break the degeneracy between strong rest-optical nebular emission and Balmer breaks for \hz\ LAEs.
 Without the medium band, the derived stellar mass could be overestimated by up to ${\sim}0.5$ dex.
 }
 \item{Almost all LAEs in our sample clearly have recent bursts of star formation. By comparing previous literature results in the SFR - $M_*$ diagram, half of our sample follow the starburst relation. Another half of them show more vigorous starburst activity. Resembling LyC leakers at \hz\ universe, the luminous LAEs function as highly efficient ionizing engines in the EoR.
 }
 \item{The LAE sample reveals non-negligible \lya\ offsets from the stellar population again. Compared to Hydrogen gas, dust content can more effectively prevent \lya\ transfer and reduce \lya\ escape for the luminous LAEs at the end of EoR. More dust content accounts for stronger suppression to the \fescLya\ and \Dd.
  }
\end{itemize}

This work exhibits strong potential to research \hz\ objects utilizing public JWST data.
Our study reveals that the luminous LAEs tend to be UV-blue, low-mass, and very young dwarf galaxies.
Their luminous \lya\ emission are generally attributed to both the bursty star-formation activities and the large escape fraction of \lya\ photons. 
As the extreme case of probable efficient ionizing engines, they are excellent subjects for JWST spectroscopic follow-up observations, including NIRSpec/IFU, to unravel the ISM conditions facilitating the escape of \lya(/ionizing) photons at the end of the EoR.

\acknowledgments

We acknowledge the support from the National Natural Science Foundation of China (grant 12403014 and 12525303), from the National Key R\&D Program of China (grant no.~2023YFA1605600), and from the Financial Program of of Beijing Academy of Science and Technology (26CA012). We also acknowledge the support from Tsinghua University Initiative Scientific Research Program, Shuimu Tsinghua Scholar Program, and New Cornerstone Science Foundation through the XPLORER PRIZE. We thank Y. Guo for the scientific discussion. We also thank the anonymous referee for the suggestions improving this work.

This work is based on the observations made with NASA/ESA/CSA James Webb Space Telescope.
The data products presented herein were retrieved from the Dawn JWST Archive (DJA). DJA is an initiative of the Cosmic Dawn Center (DAWN), which is funded by the Danish National Research Foundation under grant DNRF140. 
The DJA is built from JWST public data that originate from
the Mikulski Archive for Space Telescopes (MAST) at the Space Telescope Science Institute, which is operated by the Association of Universities for Research in Astronomy, Inc., under NASA contract NAS 5-03127 for JWST. The JWST observations are associated with the program GO-1727, GO-1837, GO-1208, GO-1210.
This work also includes data gathered with the 6.5 m Magellan Telescopes located at Las Campanas Observatory, Chile.

\facilities{Magellan: Clay (M2FS), \jwst\ (NIRCam)}
\software{BAGPIPES \citep{carnall18}, SEP \citep{barbary16}}

\bibliography{ms.bbl}

\begin{thebibliography}{}
\expandafter\ifx\csname natexlab\endcsname\relax\def\natexlab#1{#1}\fi

\bibitem[{{Arrabal Haro} {et~al.}(2023){Arrabal Haro}, {Dickinson},
  {Finkelstein}, {Fujimoto}, {Fern{\'a}ndez}, {Kartaltepe}, {Jung}, {Cole},
  {Burgarella}, {Chworowsky}, {Hutchison}, {Morales}, {Papovich}, {Simons},
  {Amor{\'\i}n}, {Backhaus}, {Bagley}, {Bisigello}, {Calabr{\`o}},
  {Castellano}, {Cleri}, {Dav{\'e}}, {Dekel}, {Ferguson}, {Fontana}, {Gawiser},
  {Giavalisco}, {Harish}, {Hathi}, {Hirschmann}, {Holwerda}, {Huertas-Company},
  {Koekemoer}, {Larson}, {Lucas}, {Mobasher}, {P{\'e}rez-Gonz{\'a}lez},
  {Pirzkal}, {Rose}, {Santini}, {Trump}, {de la Vega}, {Wang}, {Weiner},
  {Wilkins}, {Yang}, {Yung}, \& {Zavala}}]{arrabalharo23}
{Arrabal Haro}, P., {Dickinson}, M., {Finkelstein}, S.~L., {et~al.} 2023,
  \apjl, 951, L22

\bibitem[{{Asada} {et~al.}(2026){Asada}, {Fujimoto}, {Chisholm}, {Naidu},
  {Atek}, {Brammer}, {Furtak}, {Kokorev}, {Pan}, {Basu}, {Bromm},
  {Dessauges-Zavadsky}, {Hsiao}, {Jecmen}, {Korber}, {Liu}, {McKinney},
  {McQuinn}, \& {Schaerer}}]{2026arXiv260120045A}
{Asada}, Y., {Fujimoto}, S., {Chisholm}, J., {et~al.} 2026, arXiv e-prints,
  arXiv:2601.20045

\bibitem[{{Barbary}(2016)}]{barbary16}
{Barbary}, K. 2016, The Journal of Open Source Software, 1, 58

\bibitem[{{Begley} {et~al.}(2024){Begley}, {Cullen}, {McLure}, {Shapley},
  {Dunlop}, {Carnall}, {McLeod}, {Donnan}, {Hamadouche}, \&
  {Stanton}}]{begley24}
{Begley}, R., {Cullen}, F., {McLure}, R.~J., {et~al.} 2024, \mnras, 527, 4040

\bibitem[{{Bertin} \& {Arnouts}(1996)}]{ber96}
{Bertin}, E., \& {Arnouts}, S. 1996, \aaps, 117, 393

\bibitem[{{Bouwens} {et~al.}(2015){Bouwens}, {Illingworth}, {Oesch}, {Trenti},
  {Labb{\'e}}, {Bradley}, {Carollo}, {van Dokkum}, {Gonzalez}, {Holwerda},
  {Franx}, {Spitler}, {Smit}, \& {Magee}}]{bouwens15a}
{Bouwens}, R.~J., {Illingworth}, G.~D., {Oesch}, P.~A., {et~al.} 2015, \apj,
  803, 34

\bibitem[{{Bouwens} {et~al.}(2021){Bouwens}, {Oesch}, {Stefanon},
  {Illingworth}, {Labb{\'e}}, {Reddy}, {Atek}, {Montes}, {Naidu},
  {Nanayakkara}, {Nelson}, \& {Wilkins}}]{bouwens21}
{Bouwens}, R.~J., {Oesch}, P.~A., {Stefanon}, M., {et~al.} 2021, \aj, 162, 47

\bibitem[{Brammer(2023)}]{brammer23}
Brammer, G. 2023, grizli, v.1.9.11,  Zenodo, doi:10.5281/zenodo.8370018

\bibitem[{{Bunker} {et~al.}(2023){Bunker}, {Saxena}, {Cameron}, {Willott},
  {Curtis-Lake}, {Jakobsen}, {Carniani}, {Smit}, {Maiolino}, {Witstok},
  {Curti}, {D'Eugenio}, {Jones}, {Ferruit}, {Arribas}, {Charlot}, {Chevallard},
  {Giardino}, {de Graaff}, {Looser}, {L{\"u}tzgendorf}, {Maseda}, {Rawle},
  {Rix}, {Del Pino}, {Alberts}, {Egami}, {Eisenstein}, {Endsley}, {Hainline},
  {Hausen}, {Johnson}, {Rieke}, {Rieke}, {Robertson}, {Shivaei}, {Stark},
  {Sun}, {Tacchella}, {Tang}, {Williams}, {Willmer}, {Baker}, {Baum},
  {Bhatawdekar}, {Bowler}, {Boyett}, {Chen}, {Circosta}, {Helton}, {Ji},
  {Kumari}, {Lyu}, {Nelson}, {Parlanti}, {Perna}, {Sandles}, {Scholtz},
  {Suess}, {Topping}, {{\"U}bler}, {Wallace}, \& {Whitler}}]{bunker23}
{Bunker}, A.~J., {Saxena}, A., {Cameron}, A.~J., {et~al.} 2023, \aap, 677, A88

\bibitem[{{Cai} {et~al.}(2025){Cai}, {Huang}, {Liu}, {Zhao}, \&
  {Huang}}]{cai25}
{Cai}, Z., {Huang}, S., {Liu}, Y., {Zhao}, C., \& {Huang}, L. 2025, Science
  China Physics, Mechanics, and Astronomy, 68, 280403

\bibitem[{{Calzetti} {et~al.}(2000){Calzetti}, {Armus}, {Bohlin}, {Kinney},
  {Koornneef}, \& {Storchi-Bergmann}}]{calzetti00}
{Calzetti}, D., {Armus}, L., {Bohlin}, R.~C., {et~al.} 2000, \apj, 533, 682

\bibitem[{{Carnall} {et~al.}(2018){Carnall}, {McLure}, {Dunlop}, \&
  {Dav{\'e}}}]{carnall18}
{Carnall}, A.~C., {McLure}, R.~J., {Dunlop}, J.~S., \& {Dav{\'e}}, R. 2018,
  \mnras, 480, 4379

\bibitem[{{Carniani} {et~al.}(2024){Carniani}, {Hainline}, {D'Eugenio},
  {Eisenstein}, {Jakobsen}, {Witstok}, {Johnson}, {Chevallard}, {Maiolino},
  {Helton}, {Willott}, {Robertson}, {Alberts}, {Arribas}, {Baker},
  {Bhatawdekar}, {Boyett}, {Bunker}, {Cameron}, {Cargile}, {Charlot}, {Curti},
  {Curtis-Lake}, {Egami}, {Giardino}, {Isaak}, {Ji}, {Jones}, {Kumari},
  {Maseda}, {Parlanti}, {P{\'e}rez-Gonz{\'a}lez}, {Rawle}, {Rieke}, {Rieke},
  {Del Pino}, {Saxena}, {Scholtz}, {Smit}, {Sun}, {Tacchella}, {{\"U}bler},
  {Venturi}, {Williams}, \& {Willmer}}]{carniani24}
{Carniani}, S., {Hainline}, K., {D'Eugenio}, F., {et~al.} 2024, \nat, 633, 318

\bibitem[{{Casey} {et~al.}(2023){Casey}, {Kartaltepe}, {Drakos}, {Franco},
  {Harish}, {Paquereau}, {Ilbert}, {Rose}, {Cox}, {Nightingale}, {Robertson},
  {Silverman}, {Koekemoer}, {Massey}, {McCracken}, {Rhodes}, {Akins}, {Allen},
  {Amvrosiadis}, {Arango-Toro}, {Bagley}, {Bongiorno}, {Capak}, {Champagne},
  {Chartab}, {Ch{\'a}vez Ortiz}, {Chworowsky}, {Cooke}, {Cooper}, {Darvish},
  {Ding}, {Faisst}, {Finkelstein}, {Fujimoto}, {Gentile}, {Gillman}, {Gould},
  {Gozaliasl}, {Hayward}, {He}, {Hemmati}, {Hirschmann}, {Jahnke}, {Jin},
  {Khostovan}, {Kokorev}, {Lambrides}, {Laigle}, {Larson}, {Leung}, {Liu},
  {Liaudat}, {Long}, {Magdis}, {Mahler}, {Mainieri}, {Manning}, {Maraston},
  {Martin}, {McCleary}, {McKinney}, {McPartland}, {Mobasher}, {Pattnaik},
  {Renzini}, {Rich}, {Sanders}, {Sattari}, {Scognamiglio}, {Scoville}, {Sheth},
  {Shuntov}, {Sparre}, {Suzuki}, {Talia}, {Toft}, {Trakhtenbrot}, {Urry},
  {Valentino}, {Vanderhoof}, {Vardoulaki}, {Weaver}, {Whitaker}, {Wilkins},
  {Yang}, \& {Zavala}}]{casey23}
{Casey}, C.~M., {Kartaltepe}, J.~S., {Drakos}, N.~E., {et~al.} 2023, \apj, 954,
  31

\bibitem[{{Castellano} {et~al.}(2022){Castellano}, {Fontana}, {Treu},
  {Santini}, {Merlin}, {Leethochawalit}, {Trenti}, {Vanzella}, {Mestric},
  {Bonchi}, {Belfiori}, {Nonino}, {Paris}, {Polenta}, {Roberts-Borsani},
  {Boyett}, {Brada{\v{c}}}, {Calabr{\`o}}, {Glazebrook}, {Grillo}, {Mascia},
  {Mason}, {Mercurio}, {Morishita}, {Nanayakkara}, {Pentericci}, {Rosati},
  {Vulcani}, {Wang}, \& {Yang}}]{castellano22}
{Castellano}, M., {Fontana}, A., {Treu}, T., {et~al.} 2022, \apjl, 938, L15

\bibitem[{{Castellano} {et~al.}(2023){Castellano}, {Belfiori}, {Pentericci},
  {Calabr{\`o}}, {Mascia}, {Napolitano}, {Caro}, {Charlot}, {Chevallard},
  {Curtis Lake}, {Talia}, {Bongiorno}, {Fontana}, {Fynbo}, {Garilli}, {Guaita},
  {McLure}, {Merlin}, {Mignoli}, {Moresco}, {Pompei}, {Pozzetti}, {Saldana
  Lopez}, {Saxena}, {Santini}, {Schaerer}, {Schreiber}, {Shapley}, {Vanzella},
  \& {Zamorani}}]{castellano23}
{Castellano}, M., {Belfiori}, D., {Pentericci}, L., {et~al.} 2023, \aap, 675,
  A121

\bibitem[{{Castellano} {et~al.}(2024){Castellano}, {Napolitano}, {Fontana},
  {Roberts-Borsani}, {Treu}, {Vanzella}, {Zavala}, {Arrabal Haro},
  {Calabr{\`o}}, {Llerena}, {Mascia}, {Merlin}, {Paris}, {Pentericci},
  {Santini}, {Bakx}, {Bergamini}, {Cupani}, {Dickinson}, {Filippenko},
  {Glazebrook}, {Grillo}, {Kelly}, {Malkan}, {Mason}, {Morishita},
  {Nanayakkara}, {Rosati}, {Sani}, {Wang}, \& {Yoon}}]{castellano24}
{Castellano}, M., {Napolitano}, L., {Fontana}, A., {et~al.} 2024, \apj, 972,
  143

\bibitem[{{Chabrier}(2003)}]{chabrier03}
{Chabrier}, G. 2003, \pasp, 115, 763

\bibitem[{{Chen} {et~al.}(2024){Chen}, {Stark}, {Mason}, {Topping}, {Whitler},
  {Tang}, {Endsley}, \& {Charlot}}]{chenzy24}
{Chen}, Z., {Stark}, D.~P., {Mason}, C., {et~al.} 2024, \mnras, 528, 7052

\bibitem[{{Choustikov} {et~al.}(2024){Choustikov}, {Katz}, {Saxena}, {Garel},
  {Devriendt}, {Slyz}, {Kimm}, {Blaizot}, \& {Rosdahl}}]{choustikov24}
{Choustikov}, N., {Katz}, H., {Saxena}, A., {et~al.} 2024, \mnras, 532, 2463

\bibitem[{{Clarke} {et~al.}(2024){Clarke}, {Shapley}, {Sanders}, {Topping},
  {Brammer}, {Bento}, {Reddy}, \& {Kehoe}}]{clarke24}
{Clarke}, L., {Shapley}, A.~E., {Sanders}, R.~L., {et~al.} 2024, \apj, 977, 133

\bibitem[{{Curti} {et~al.}(2024){Curti}, {Maiolino}, {Curtis-Lake},
  {Chevallard}, {Carniani}, {D'Eugenio}, {Looser}, {Scholtz}, {Charlot},
  {Cameron}, {{\"U}bler}, {Witstok}, {Boyett}, {Laseter}, {Sandles}, {Arribas},
  {Bunker}, {Giardino}, {Maseda}, {Rawle}, {Rodr{\'\i}guez Del Pino}, {Smit},
  {Willott}, {Eisenstein}, {Hausen}, {Johnson}, {Rieke}, {Robertson},
  {Tacchella}, {Williams}, {Willmer}, {Baker}, {Bhatawdekar}, {Egami},
  {Helton}, {Ji}, {Kumari}, {Perna}, {Shivaei}, \& {Sun}}]{curti24}
{Curti}, M., {Maiolino}, R., {Curtis-Lake}, E., {et~al.} 2024, \aap, 684, A75

\bibitem[{Dijkstra(2014)}]{dijkstra14}
Dijkstra, M. 2014, PASA, 31, e040

\bibitem[{{Dijkstra} {et~al.}(2016){Dijkstra}, {Gronke}, \&
  {Venkatesan}}]{dijkstra16}
{Dijkstra}, M., {Gronke}, M., \& {Venkatesan}, A. 2016, \apj, 828, 71

\bibitem[{{Dijkstra} {et~al.}(2011){Dijkstra}, {Mesinger}, \&
  {Wyithe}}]{dijkstra11}
{Dijkstra}, M., {Mesinger}, A., \& {Wyithe}, J. S.~B. 2011, \mnras, 414, 2139

\bibitem[{{Donnan} {et~al.}(2023){Donnan}, {McLeod}, {Dunlop}, {McLure},
  {Carnall}, {Begley}, {Cullen}, {Hamadouche}, {Bowler}, {Magee}, {McCracken},
  {Milvang-Jensen}, {Moneti}, \& {Targett}}]{donnan23}
{Donnan}, C.~T., {McLeod}, D.~J., {Dunlop}, J.~S., {et~al.} 2023, \mnras, 518,
  6011

\bibitem[{{Douglas} {et~al.}(2009){Douglas}, {Bremer}, {Stanway}, {Lehnert}, \&
  {Clowe}}]{douglas09}
{Douglas}, L.~S., {Bremer}, M.~N., {Stanway}, E.~R., {Lehnert}, M.~D., \&
  {Clowe}, D. 2009, \mnras, 400, 561

\bibitem[{{Dunlop} {et~al.}(2021){Dunlop}, {Abraham}, {Ashby}, {Bagley},
  {Best}, {Bongiorno}, {Bouwens}, {Bowler}, {Brammer}, {Bremer}, {Calabro'},
  {Carnall}, {Castellano}, {Cirasuolo}, {Conselice}, {Cullen}, {Dave}, {Dayal},
  {Dekel}, {Dickinson}, {Duncan}, {Elbaz}, {Ellis}, {Ferguson}, {Ferrara},
  {Finkelstein}, {Fontana}, {Furlanetto}, {Fynbo}, {Gallerani}, {Gardner},
  {Giavalisco}, {Grazian}, {Grogin}, {Harikane}, {Hopkins}, {Ilbert},
  {Illingworth}, {Juneau}, {Jung}, {Kartaltepe}, {Kassin}, {Kauffmann},
  {Khochfar}, {Kirkpatrick}, {Kocevski}, {Koekemoer}, {Labbe}, {Laporte},
  {Larson}, {Lucas}, {Magee}, {Mason}, {McCracken}, {McLeod}, {McLure},
  {Merlin}, {Mesinger}, {Milvang-Jensen}, {Newman}, {Oesch}, {Ouchi},
  {Pacifici}, {Papovich}, {Peacock}, {Peeples}, {Pentericci}, {Perez-Gonzalez},
  {Pirzkal}, {Pope}, {Pye}, {Reddy}, {Robertson}, {Salvato}, {Santini},
  {Schaerer}, {Shapley}, {Simons}, {Smit}, {Smith}, {Snyder}, {Somerville},
  {Stanway}, {Stefanon}, {Tasca}, {Tikkanen}, {Tresse}, {Trump}, {Whitaker},
  {Wilkins}, {Wright}, {Wyithe}, {van Dokkum}, \& {van der Werf}}]{primer}
{Dunlop}, J.~S., {Abraham}, R.~G., {Ashby}, M. L.~N., {et~al.} 2021, {PRIMER:
  Public Release IMaging for Extragalactic Research}, JWST Proposal. Cycle 1,
  ID. \#1837, ,

\bibitem[{{Eisenstein} {et~al.}(2026){Eisenstein}, {Willott}, {Alberts},
  {Arribas}, {Bonaventura}, {Bunker}, {Cameron}, {Carniani}, {Charlot},
  {Curtis-Lake}, {D'Eugenio}, {Ferruit}, {Giardino}, {Hainline}, {Hausen},
  {Jakobsen}, {Johnson}, {Maiolino}, {Rauscher}, {Rieke}, {Rieke}, {Rix},
  {Robertson}, {Stark}, {Tacchella}, {Williams}, {Willmer}, {Baker}, {Baum},
  {Bhatawdekar}, {Boyett}, {Chen}, {Chevallard}, {Circosta}, {Curti},
  {Danhaive}, {DeCoursey}, {Endsley}, {de Graaff}, {Dressler}, {Egami},
  {Helton}, {Hviding}, {Ji}, {Jones}, {Kumari}, {L{\"u}tzgendorf}, {Laseter},
  {Looser}, {Lyu}, {Maseda}, {Nelson}, {Parlanti}, {Perna}, {Pusk{\'a}s},
  {Rawle}, {Rodr{\'\i}guez Del Pino}, {Rujopakarn}, {Sandles}, {Saxena},
  {Scholtz}, {Sharpe}, {Shivaei}, {Silcock}, {Simmonds}, {Skarbinski}, {Smit},
  {Stone}, {Suess}, {Sun}, {Tang}, {Topping}, {{\"U}bler}, {Villanueva},
  {Wallace}, {Whitler}, {Witstok}, \& {Woodrum}}]{jades26}
{Eisenstein}, D.~J., {Willott}, C., {Alberts}, S., {et~al.} 2026, \apjs, 283, 6

\bibitem[{{Ellis} {et~al.}(2013){Ellis}, {McLure}, {Dunlop}, {Robertson},
  {Ono}, {Schenker}, {Koekemoer}, {Bowler}, {Ouchi}, {Rogers}, {Curtis-Lake},
  {Schneider}, {Charlot}, {Stark}, {Furlanetto}, \& {Cirasuolo}}]{ellis13}
{Ellis}, R.~S., {McLure}, R.~J., {Dunlop}, J.~S., {et~al.} 2013, \apjl, 763, L7

\bibitem[{{Faisst} {et~al.}(2019){Faisst}, {Capak}, {Emami}, {Tacchella}, \&
  {Larson}}]{faisst19}
{Faisst}, A.~L., {Capak}, P.~L., {Emami}, N., {Tacchella}, S., \& {Larson},
  K.~L. 2019, \apj, 884, 133

\bibitem[{{Finkelstein} {et~al.}(2022){Finkelstein}, {Bagley}, {Arrabal Haro},
  {Dickinson}, {Ferguson}, {Kartaltepe}, {Papovich}, {Burgarella}, {Kocevski},
  {Huertas-Company}, {Iyer}, {Koekemoer}, {Larson}, {P{\'e}rez-Gonz{\'a}lez},
  {Rose}, {Tacchella}, {Wilkins}, {Chworowsky}, {Medrano}, {Morales},
  {Somerville}, {Yung}, {Fontana}, {Giavalisco}, {Grazian}, {Grogin}, {Kewley},
  {Kirkpatrick}, {Kurczynski}, {Lotz}, {Pentericci}, {Pirzkal}, {Ravindranath},
  {Ryan}, {Trump}, {Yang}, {Almaini}, {Amor{\'\i}n}, {Annunziatella},
  {Backhaus}, {Barro}, {Behroozi}, {Bell}, {Bhatawdekar}, {Bisigello}, {Bromm},
  {Buat}, {Buitrago}, {Calabr{\`o}}, {Casey}, {Castellano}, {Ch{\'a}vez Ortiz},
  {Ciesla}, {Cleri}, {Cohen}, {Cole}, {Cooke}, {Cooper}, {Cooray}, {Costantin},
  {Cox}, {Croton}, {Daddi}, {Dav{\'e}}, {de La Vega}, {Dekel}, {Elbaz},
  {Estrada-Carpenter}, {Faber}, {Fern{\'a}ndez}, {Finkelstein}, {Freundlich},
  {Fujimoto}, {Garc{\'\i}a-Argum{\'a}nez}, {Gardner}, {Gawiser},
  {G{\'o}mez-Guijarro}, {Guo}, {Hamblin}, {Hamilton}, {Hathi}, {Holwerda},
  {Hirschmann}, {Hutchison}, {Jaskot}, {Jha}, {Jogee}, {Juneau}, {Jung},
  {Kassin}, {Le Bail}, {Leung}, {Lucas}, {Magnelli}, {Mantha}, {Matharu},
  {McGrath}, {McIntosh}, {Merlin}, {Mobasher}, {Newman}, {Nicholls}, {Pandya},
  {Rafelski}, {Ronayne}, {Santini}, {Seill{\'e}}, {Shah}, {Shen}, {Simons},
  {Snyder}, {Stanway}, {Straughn}, {Teplitz}, {Vanderhoof}, {Vega-Ferrero},
  {Wang}, {Weiner}, {Willmer}, {Wuyts}, {Zavala}, \& {Ceers
  Team}}]{finkelstein22}
{Finkelstein}, S.~L., {Bagley}, M.~B., {Arrabal Haro}, P., {et~al.} 2022,
  \apjl, 940, L55

\bibitem[{{Finkelstein} {et~al.}(2024){Finkelstein}, {Leung}, {Bagley},
  {Dickinson}, {Ferguson}, {Papovich}, {Akins}, {Arrabal Haro}, {Dav{\'e}},
  {Dekel}, {Kartaltepe}, {Kocevski}, {Koekemoer}, {Pirzkal}, {Somerville},
  {Yung}, {Amor{\'\i}n}, {Backhaus}, {Behroozi}, {Bisigello}, {Bromm}, {Casey},
  {Ch{\'a}vez Ortiz}, {Cheng}, {Chworowsky}, {Cleri}, {Cooper}, {Davis}, {de la
  Vega}, {Elbaz}, {Franco}, {Fontana}, {Fujimoto}, {Giavalisco}, {Grogin},
  {Holwerda}, {Huertas-Company}, {Hirschmann}, {Iyer}, {Jogee}, {Jung},
  {Larson}, {Lucas}, {Mobasher}, {Morales}, {Morley}, {Mukherjee},
  {P{\'e}rez-Gonz{\'a}lez}, {Ravindranath}, {Rodighiero}, {Rowland},
  {Tacchella}, {Taylor}, {Trump}, \& {Wilkins}}]{finkelstein24}
{Finkelstein}, S.~L., {Leung}, G. C.~K., {Bagley}, M.~B., {et~al.} 2024, \apjl,
  969, L2

\bibitem[{{Firestone} {et~al.}(2025){Firestone}, {Gawiser}, {Iyer}, {Lee},
  {Ramakrishnan}, {Valdes}, {Park}, {Yang}, {Alavi}, {Ciardullo}, {Grogin},
  {Gronwall}, {Guaita}, {Hong}, {Hwang}, {Im}, {Jeong}, {Kim}, {Koekemoer},
  {Kumar}, {Lee}, {Mehta}, {Nagaraj}, {Nantais}, {Prichard}, {Rafelski},
  {Song}, {Sunnquist}, {Teplitz}, \& {Wang}}]{firestone25}
{Firestone}, N.~M., {Gawiser}, E., {Iyer}, K.~G., {et~al.} 2025, \apjl, 986, L8

\bibitem[{{Franco} {et~al.}(2024){Franco}, {Akins}, {Casey}, {Finkelstein},
  {Shuntov}, {Chworowsky}, {Faisst}, {Fujimoto}, {Ilbert}, {Koekemoer}, {Liu},
  {Lovell}, {Maraston}, {McCracken}, {McKinney}, {Robertson}, {Bagley},
  {Champagne}, {Cooper}, {Ding}, {Drakos}, {Enia}, {Gillman}, {Gozaliasl},
  {Harish}, {Hayward}, {Hirschmann}, {Jin}, {Kartaltepe}, {Kokorev}, {Laigle},
  {Long}, {Magdis}, {Mahler}, {Martin}, {Massey}, {Mobasher}, {Paquereau},
  {Renzini}, {Rhodes}, {Rich}, {Sheth}, {Silverman}, {Sparre}, {Talia},
  {Trakhtenbrot}, {Valentino}, {Vijayan}, {Wilkins}, {Yang}, \&
  {Zavala}}]{franco24}
{Franco}, M., {Akins}, H.~B., {Casey}, C.~M., {et~al.} 2024, \apj, 973, 23

\bibitem[{{Fu} {et~al.}(2024){Fu}, {Jiang}, {Ning}, {Liu}, \& {Pan}}]{fusq24}
{Fu}, S., {Jiang}, L., {Ning}, Y., {Liu}, W., \& {Pan}, Z. 2024, \apj, 963, 51

\bibitem[{{Gazagnes} {et~al.}(2020){Gazagnes}, {Chisholm}, {Schaerer},
  {Verhamme}, \& {Izotov}}]{gazagnes20}
{Gazagnes}, S., {Chisholm}, J., {Schaerer}, D., {Verhamme}, A., \& {Izotov}, Y.
  2020, \aap, 639, A85

\bibitem[{{Giavalisco}(2002)}]{giavalisco02}
{Giavalisco}, M. 2002, \araa, 40, 579

\bibitem[{{Goovaerts} {et~al.}(2024){Goovaerts}, {Pello}, {Burgarella}, {Thai},
  {Richard}, {Claeyssens}, {Tuan-Anh}, {Arango-Toro}, {Boogaard}, {Contini},
  {Guo}, {Langan}, {Laporte}, \& {Maseda}}]{goovaerts24}
{Goovaerts}, I., {Pello}, R., {Burgarella}, D., {et~al.} 2024, \aap, 683, A184

\bibitem[{{Greene} {et~al.}(2024){Greene}, {Labbe}, {Goulding}, {Furtak},
  {Chemerynska}, {Kokorev}, {Dayal}, {Volonteri}, {Williams}, {Wang}, {Setton},
  {Burgasser}, {Bezanson}, {Atek}, {Brammer}, {Cutler}, {Feldmann}, {Fujimoto},
  {Glazebrook}, {de Graaff}, {Khullar}, {Leja}, {Marchesini}, {Maseda},
  {Matthee}, {Miller}, {Naidu}, {Nanayakkara}, {Oesch}, {Pan}, {Papovich},
  {Price}, {van Dokkum}, {Weaver}, {Whitaker}, \& {Zitrin}}]{greene24}
{Greene}, J.~E., {Labbe}, I., {Goulding}, A.~D., {et~al.} 2024, \apj, 964, 39

\bibitem[{{Hayes}(2015)}]{hayes15}
{Hayes}, M. 2015, \pasa, 32, e027

\bibitem[{{Heintz} {et~al.}(2025){Heintz}, {Brammer}, {Watson}, {Oesch},
  {Keating}, {Hayes}, {Abdurro'uf}, {Arellano-C{\'o}rdova}, {Carnall},
  {Christiansen}, {Cullen}, {Dav{\'e}}, {Dayal}, {Ferrara}, {Finlator},
  {Fynbo}, {Flury}, {Gelli}, {Gillman}, {Gottumukkala}, {Gould}, {Greve},
  {Hardin}, {Hsiao}, {Hutter}, {Jakobsson}, {Killi}, {Khosravaninezhad},
  {Laursen}, {Lee}, {Magdis}, {Matthee}, {Naidu}, {Narayanan}, {Pollock},
  {Prescott}, {Rusakov}, {Shuntov}, {Sneppen}, {Smit}, {Tanvir}, {Terp},
  {Toft}, {Valentino}, {Vijayan}, {Weaver}, {Wise}, \& {Witstok}}]{heintz25}
{Heintz}, K.~E., {Brammer}, G.~B., {Watson}, D., {et~al.} 2025, \aap, 693, A60

\bibitem[{{Hern{\'a}n-Caballero} {et~al.}(2017){Hern{\'a}n-Caballero},
  {P{\'e}rez-Gonz{\'a}lez}, {Diego}, {Lagattuta}, {Richard}, {Schaerer},
  {Alonso-Herrero}, {Marino}, {Sklias}, {Alcalde Pampliega}, {Cava},
  {Conselice}, {Dannerbauer}, {Dom{\'\i}nguez-S{\'a}nchez}, {Eliche-Moral},
  {Esquej}, {Huertas-Company}, {Marques-Chaves}, {P{\'e}rez-Fournon}, {Rawle},
  {Rodr{\'\i}guez Espinosa}, {Rosa Gonz{\'a}lez}, \& {Rujopakarn}}]{hc17}
{Hern{\'a}n-Caballero}, A., {P{\'e}rez-Gonz{\'a}lez}, P.~G., {Diego}, J.~M.,
  {et~al.} 2017, \apj, 849, 82

\bibitem[{{Hu} {et~al.}(2010){Hu}, {Cowie}, {Barger}, {Capak}, {Kakazu}, \&
  {Trouille}}]{hu10}
{Hu}, E.~M., {Cowie}, L.~L., {Barger}, A.~J., {et~al.} 2010, \apj, 725, 394

\bibitem[{{Hu} {et~al.}(2002){Hu}, {Cowie}, {McMahon}, {Capak}, {Iwamuro},
  {Kneib}, {Maihara}, \& {Motohara}}]{hu02}
{Hu}, E.~M., {Cowie}, L.~L., {McMahon}, R.~G., {et~al.} 2002, \apj, 568, L75

\bibitem[{{Hu} \& {McMahon}(1996)}]{hu96}
{Hu}, E.~M., \& {McMahon}, R.~G. 1996, \nat, 382, 231

\bibitem[{{Iani} {et~al.}(2024){Iani}, {Caputi}, {Rinaldi}, {Annunziatella},
  {Boogaard}, {{\"O}stlin}, {Costantin}, {Gillman}, {P{\'e}rez-Gonz{\'a}lez},
  {Colina}, {Greve}, {Wright}, {Alonso-Herrero}, {{\'A}lvarez-M{\'a}rquez},
  {Bik}, {Bosman}, {Crespo G{\'o}mez}, {Eckart}, {Hjorth}, {Jermann},
  {Labiano}, {Langeroodi}, {Melinder}, {Moutard}, {Pei{\ss}ker}, {Pye},
  {Tikkanen}, {van der Werf}, {Walter}, {Henning}, {Lagage}, \& {van
  Dishoeck}}]{iani24}
{Iani}, E., {Caputi}, K.~I., {Rinaldi}, P., {et~al.} 2024, \apj, 963, 97

\bibitem[{{Ismail} {et~al.}(2026){Ismail}, {Kraljic}, {B{\'e}thermin},
  {Kapoor}, {Renaud}, {Accard}, {Freundlich}, {Han}, {Jang}, {Jeon}, {Kimm},
  {Rhee}, \& {Yi}}]{2026arXiv260105916I}
{Ismail}, D., {Kraljic}, K., {B{\'e}thermin}, M., {et~al.} 2026, arXiv
  e-prints, arXiv:2601.05916

\bibitem[{{Iye} {et~al.}(2006){Iye}, {Ota}, {Kashikawa}, {Furusawa},
  {Hashimoto}, {Hattori}, {Matsuda}, {Morokuma}, {Ouchi}, \&
  {Shimasaku}}]{iye06}
{Iye}, M., {Ota}, K., {Kashikawa}, N., {et~al.} 2006, \nat, 443, 186

\bibitem[{{Izotov} {et~al.}(2020){Izotov}, {Schaerer}, {Worseck}, {Verhamme},
  {Guseva}, {Thuan}, {Orlitov{\'a}}, \& {Fricke}}]{izotov20}
{Izotov}, Y.~I., {Schaerer}, D., {Worseck}, G., {et~al.} 2020, \mnras, 491, 468

\bibitem[{{Izotov} {et~al.}(2024){Izotov}, {Thuan}, {Guseva}, {Schaerer},
  {Worseck}, \& {Verhamme}}]{izotov24}
{Izotov}, Y.~I., {Thuan}, T.~X., {Guseva}, N.~G., {et~al.} 2024, \mnras, 527,
  281

\bibitem[{{Jecmen} {et~al.}(2026){Jecmen}, {Chisholm}, {Atek}, {Kokorev},
  {Endsley}, {Chemerynska}, {Furtak}, {Pan}, {Fujimoto}, {Naidu}, {Mu{\~n}oz},
  {Adamo}, {Asada}, {Basu}, {Berg}, {Blaizot}, {Dessauges-Zavadsky},
  {Giovinazzo}, {Hsiao}, {Katz}, {Korber}, {McKinney}, {McQuinn}, {Oesch}, \&
  {Schaerer}}]{2026arXiv260119995J}
{Jecmen}, M.~C., {Chisholm}, J., {Atek}, H., {et~al.} 2026, arXiv e-prints,
  arXiv:2601.19995

\bibitem[{{Jiang} {et~al.}(2024){Jiang}, {Wang}, {Cheng}, {Kong}, {Zhou},
  {Meng}, {He}, {Jones}, \& {Boyett}}]{jianghc24}
{Jiang}, H., {Wang}, X., {Cheng}, C., {et~al.} 2024, \apj, 972, 121

\bibitem[{{Jiang} {et~al.}(2020){Jiang}, {Cohen}, {Windhorst}, {Egami},
  {Finlator}, {Schaerer}, \& {Sun}}]{jiang20a}
{Jiang}, L., {Cohen}, S.~H., {Windhorst}, R.~A., {et~al.} 2020, \apj, 889, 90

\bibitem[{{Jiang} {et~al.}(2013){Jiang}, {Egami}, {Mechtley}, {Fan}, {Cohen},
  {Windhorst}, {Dav{\'e}}, {Finlator}, {Kashikawa}, {Ouchi}, \&
  {Shimasaku}}]{jiang13a}
{Jiang}, L., {Egami}, E., {Mechtley}, M., {et~al.} 2013, \apj, 772, 99

\bibitem[{{Jiang} {et~al.}(2016){Jiang}, {Finlator}, {Cohen}, {Egami},
  {Windhorst}, {Fan}, {Dav{\'e}}, {Kashikawa}, {Mechtley}, {Ouchi},
  {Shimasaku}, \& {Cl{\'e}ment}}]{jiang16a}
{Jiang}, L., {Finlator}, K., {Cohen}, S.~H., {et~al.} 2016, \apj, 816, 16

\bibitem[{{Jiang} {et~al.}(2017){Jiang}, {Shen}, {Bian}, {Zheng}, {Wu},
  {Oyarz{\'u}n}, {Blanc}, {Fan}, {Ho}, {Infante}, {Wang}, {Wu}, {Mateo},
  {Bailey}, {Crane}, {Olszewski}, {Shectman}, {Thompson}, \&
  {Walker}}]{jiang17}
{Jiang}, L., {Shen}, Y., {Bian}, F., {et~al.} 2017, \apj, 846, 134

\bibitem[{{Jones} {et~al.}(2024){Jones}, {Bunker}, {Saxena}, {Witstok},
  {Stark}, {Arribas}, {Baker}, {Bhatawdekar}, {Bowler}, {Boyett}, {Cameron},
  {Carniani}, {Charlot}, {Chevallard}, {Curti}, {Curtis-Lake}, {Eisenstein},
  {Hainline}, {Hausen}, {Ji}, {Johnson}, {Kumari}, {Looser}, {Maiolino},
  {Maseda}, {Parlanti}, {Rix}, {Robertson}, {Sandles}, {Scholtz}, {Smit},
  {Tacchella}, {{\"U}bler}, {Williams}, \& {Willott}}]{jones24}
{Jones}, G.~C., {Bunker}, A.~J., {Saxena}, A., {et~al.} 2024, \aap, 683, A238

\bibitem[{{Kartaltepe} {et~al.}(2021){Kartaltepe}, {Casey}, {Bagley},
  {Bongiorno}, {Capak}, {Champagne}, {Cooke}, {Cooper}, {Darvish}, {Davidzon},
  {Drakos}, {Drew}, {Faisst}, {Finkelstein}, {Hayward}, {Hemmati},
  {Hirschmann}, {Ilbert}, {Jahnke}, {Koekemoer}, {Liu}, {Long}, {Magdis},
  {Manning}, {Maraston}, {Martin}, {Massey}, {McCleary}, {McCracken},
  {Nayyeri}, {Renzini}, {Rhodes}, {Rich}, {Robertson}, {Rose}, {Sanders},
  {Scarlata}, {Scoville}, {Sheth}, {Silverman}, {Sparre}, {Talia}, {Toft},
  {Trakhtenbrot}, {Vanderhoof}, {Vardoulaki}, {Whitaker}, {Wilkins}, \&
  {Zavala}}]{cosmosWeb}
{Kartaltepe}, J., {Casey}, C.~M., {Bagley}, M., {et~al.} 2021, {COSMOS-Webb:
  The Webb Cosmic Origins Survey}, JWST Proposal. Cycle 1, ID. \#1727, ,

\bibitem[{{Kashikawa} {et~al.}(2011){Kashikawa}, {Shimasaku}, {Matsuda},
  {Egami}, {Jiang}, {Nagao}, {Ouchi}, {Malkan}, {Hattori}, {Ota}, {Taniguchi},
  {Okamura}, {Ly}, {Iye}, {Furusawa}, {Shioya}, {Shibuya}, {Ishizaki}, \&
  {Toshikawa}}]{kashikawa11}
{Kashikawa}, N., {Shimasaku}, K., {Matsuda}, Y., {et~al.} 2011, \apj, 734, 119

\bibitem[{{Kennicutt} {et~al.}(1994){Kennicutt}, {Tamblyn}, \&
  {Congdon}}]{kennicutt94}
{Kennicutt}, Robert~C., J., {Tamblyn}, P., \& {Congdon}, C.~E. 1994, \apj, 435,
  22

\bibitem[{{Kennicutt}(1998)}]{kennicutt98b}
{Kennicutt}, Jr., R.~C. 1998, \apj, 498, 541

\bibitem[{{Kikuta} {et~al.}(2023){Kikuta}, {Ouchi}, {Shibuya}, {Liang},
  {Umeda}, {Matsumoto}, {Shimasaku}, {Harikane}, {Ono}, {Inoue}, {Yamanaka},
  {Kusakabe}, {Momose}, {Kashikawa}, {Matsuda}, \& {Lee}}]{kikuta23}
{Kikuta}, S., {Ouchi}, M., {Shibuya}, T., {et~al.} 2023, \apjs, 268, 24

\bibitem[{{Kim} {et~al.}(2021){Kim}, {Malhotra}, {Rhoads}, \& {Yang}}]{kimk21}
{Kim}, K.~J., {Malhotra}, S., {Rhoads}, J.~E., \& {Yang}, H. 2021, \apj, 914, 2

\bibitem[{{Kong} {et~al.}(2026){Kong}, {Jones}, {Drakos}, {Malhotra}, {Iyer},
  {Lemaux}, {Naidu}, {de Boer}, {Chambers}, {Fairlamb}, {Hoogendam}, {Huber},
  {Lin}, {Lowe}, {Magnier}, {M{\'\i}nguez}, {Paek}, {Schultz}, \&
  {Wainscoat}}]{2026arXiv260211261K}
{Kong}, M.~Y., {Jones}, D.~O., {Drakos}, N.~E., {et~al.} 2026, arXiv e-prints,
  arXiv:2602.11261

\bibitem[{{Kramarenko} {et~al.}(2025){Kramarenko}, {Rosdahl}, {Blaizot},
  {Matthee}, {Katz}, \& {Di Cesare}}]{2025arXiv250905403K}
{Kramarenko}, I.~G., {Rosdahl}, J., {Blaizot}, J., {et~al.} 2025, arXiv
  e-prints, arXiv:2509.05403

\bibitem[{{Lange}(2023)}]{lange23}
{Lange}, J.~U. 2023, \mnras, 525, 3181

\bibitem[{{Leitherer} \& {Heckman}(1995)}]{lh95}
{Leitherer}, C., \& {Heckman}, T.~M. 1995, \apjs, 96, 9

\bibitem[{{Li} {et~al.}(2026){Li}, {Conselice}, {Austin}, {Harvey}, {Adams},
  {Rusakov}, \& {Westcott}}]{liq26}
{Li}, Q., {Conselice}, C.~J., {Austin}, D., {et~al.} 2026, \mnras, 547, stag311

\bibitem[{{Llerena} {et~al.}(2025){Llerena}, {Pentericci}, {Amor{\'\i}n},
  {Ferrara}, {Dickinson}, {Arevalo}, {Calabr{\`o}}, {Napolitano}, {Mascia},
  {Arrabal Haro}, {Begley}, {Cleri}, {Davis}, {Hu}, {Kartaltepe}, {Koekemoer},
  {Lucas}, {McGrath}, {McLeod}, {Papovich}, {Stanton}, {Taylor}, {Tripodi},
  {Wang}, \& {Yung}}]{2025arXiv251025647L}
{Llerena}, M., {Pentericci}, L., {Amor{\'\i}n}, R., {et~al.} 2025, arXiv
  e-prints, arXiv:2510.25647

\bibitem[{{Lyu} {et~al.}(2025){Lyu}, {Wang}, {Wang}, {Jia}, {Song}, {Chen},
  {Chen}, {Yu}, {Ma}, {Wang}, {Wang}, \& {Kong}}]{lyuc25}
{Lyu}, C., {Wang}, E., {Wang}, J., {et~al.} 2025, \apjl, 988, L72

\bibitem[{{Madau} {et~al.}(1998){Madau}, {Pozzetti}, \& {Dickinson}}]{madau98}
{Madau}, P., {Pozzetti}, L., \& {Dickinson}, M. 1998, \apj, 498, 106

\bibitem[{{Mao} {et~al.}(2007){Mao}, {Lapi}, {Granato}, {de Zotti}, \&
  {Danese}}]{maoj07}
{Mao}, J., {Lapi}, A., {Granato}, G.~L., {de Zotti}, G., \& {Danese}, L. 2007,
  \apj, 667, 655

\bibitem[{{Markov} {et~al.}(2025){Markov}, {Brada{\v{c}}}, {Estrada-Carpenter},
  {Desprez}, {Rihtar{\v{s}}i{\v{c}}}, {Jude{\v{z}}}, {Tripodi}, {Sawicki},
  {Noirot}, {Martis}, {Willott}, {Sarrouh}, {Withers}, {Muzzin}, {Asada},
  {Gallerani}, {Ferrara}, {Goovaerts}, {Harshan}, \&
  {Fujimoto}}]{2025arXiv251213778M}
{Markov}, V., {Brada{\v{c}}}, M., {Estrada-Carpenter}, V., {et~al.} 2025, arXiv
  e-prints, arXiv:2512.13778

\bibitem[{{Marques-Chaves} {et~al.}(2026){Marques-Chaves},
  {{\'A}lvarez-M{\'a}rquez}, {Colina}, {Kendrew}, {Abdurro'uf},
  {Blanco-Prieto}, {Boogaard}, {Castellano}, {Caputi}, {Crespo-Gomez},
  {Fontana}, {Fudamoto}, {Fujimoto}, {Garc{\'\i}a-Mar{\'\i}n}, {Harikane},
  {Harish}, {Hashimoto}, {Hsiao}, {Iani}, {Inoue}, {Langeroodi}, {Lin},
  {Melinder}, {Napolitano}, {Ostlin}, {P{\'e}rez-Gonz{\'a}lez},
  {Prieto-Jim{\'e}nez}, {Rinaldi}, {Rodr{\'\i}guez Del Pino}, {Santini},
  {Sugahara}, {Varo-O'ferral}, {Wright}, \& {Zavala}}]{2026arXiv260202322M}
{Marques-Chaves}, R., {{\'A}lvarez-M{\'a}rquez}, J., {Colina}, L., {et~al.}
  2026, arXiv e-prints, arXiv:2602.02322

\bibitem[{{Martin} {et~al.}(2008){Martin}, {Sawicki}, {Dressler}, \&
  {McCarthy}}]{martin08}
{Martin}, C.~L., {Sawicki}, M., {Dressler}, A., \& {McCarthy}, P. 2008, \apj,
  679, 942

\bibitem[{{McLeod} {et~al.}(2016){McLeod}, {McLure}, \& {Dunlop}}]{mcLeod16}
{McLeod}, D.~J., {McLure}, R.~J., \& {Dunlop}, J.~S. 2016, \mnras, 459, 3812

\bibitem[{{Napolitano} {et~al.}(2024){Napolitano}, {Pentericci}, {Santini},
  {Calabr{\`o}}, {Mascia}, {Llerena}, {Castellano}, {Dickinson}, {Finkelstein},
  {Amor{\'\i}n}, {Arrabal Haro}, {Bagley}, {Bhatawdekar}, {Cleri}, {Davis},
  {Gardner}, {Gawiser}, {Giavalisco}, {Hathi}, {Holwerda}, {Hu}, {Jung},
  {Kartaltepe}, {Koekemoer}, {Larson}, {Merlin}, {Mobasher}, {Papovich},
  {Park}, {Pirzkal}, {Trump}, {Wilkins}, \& {Yung}}]{napolitano24}
{Napolitano}, L., {Pentericci}, L., {Santini}, P., {et~al.} 2024, \aap, 688,
  A106

\bibitem[{{Navarro-Carrera} {et~al.}(2025){Navarro-Carrera}, {Caputi}, {Iani},
  {Rinaldi}, {Kokorev}, \& {Kerutt}}]{navarro-carrera25}
{Navarro-Carrera}, R., {Caputi}, K.~I., {Iani}, E., {et~al.} 2025, \apj, 993,
  194

\bibitem[{{Ning} {et~al.}(2023){Ning}, {Cai}, {Jiang}, {Lin}, {Fu}, \&
  {Spinoso}}]{ning23}
{Ning}, Y., {Cai}, Z., {Jiang}, L., {et~al.} 2023, \apjl, 944, L1

\bibitem[{{Ning} {et~al.}(2022){Ning}, {Jiang}, {Zheng}, \& {Wu}}]{ning22}
{Ning}, Y., {Jiang}, L., {Zheng}, Z.-Y., \& {Wu}, J. 2022, \apj, 926, 230

\bibitem[{{Ning} {et~al.}(2020){Ning}, {Jiang}, {Zheng}, {Wu}, {Bian}, {Egami},
  {Fan}, {Ho}, {Shen}, {Wang}, \& {Wu}}]{ning20}
{Ning}, Y., {Jiang}, L., {Zheng}, Z.-Y., {et~al.} 2020, \apj, 903, 4

\bibitem[{{Ning} {et~al.}(2024){Ning}, {Cai}, {Lin}, {Zheng}, {Feng}, {Li},
  {Li}, {Spinoso}, {Wu}, \& {Zhang}}]{ning24}
{Ning}, Y., {Cai}, Z., {Lin}, X., {et~al.} 2024, \apjl, 963, L38

\bibitem[{{Oesch} {et~al.}(2018){Oesch}, {Bouwens}, {Illingworth}, {Labb{\'e}},
  \& {Stefanon}}]{oesch18}
{Oesch}, P.~A., {Bouwens}, R.~J., {Illingworth}, G.~D., {Labb{\'e}}, I., \&
  {Stefanon}, M. 2018, \apj, 855, 105

\bibitem[{{Osterbrock} \& {Ferland}(2006)}]{agn2}
{Osterbrock}, D.~E., \& {Ferland}, G.~J. 2006, {Astrophysics of gaseous nebulae
  and active galactic nuclei}

\bibitem[{{Ouchi}(2019)}]{ouchi20}
{Ouchi}, M. 2019, Saas-Fee Advanced Course, 46, 189

\bibitem[{{Partridge} \& {Peebles}(1967)}]{pp67}
{Partridge}, R.~B., \& {Peebles}, P.~J.~E. 1967, \apj, 147, 868

\bibitem[{{Pentericci} {et~al.}(2018){Pentericci}, {Vanzella}, {Castellano},
  {Fontana}, {De Barros}, {Grazian}, {Marchi}, {Bradac}, {Conselice},
  {Cristiani}, {Dickinson}, {Finkelstein}, {Giallongo}, {Guaita}, {Koekemoer},
  {Maiolino}, {Santini}, \& {Tilvi}}]{pentericci18b}
{Pentericci}, L., {Vanzella}, E., {Castellano}, M., {et~al.} 2018, \aap, 619,
  A147

\bibitem[{{Prieto-Lyon} {et~al.}(2025){Prieto-Lyon}, {Mason}, {Strait},
  {Brammer}, {Naidu}, {Meyer}, {Oesch}, {Tacchella}, {Covelo-Paz},
  {Giovinazzo}, \& {Xiao}}]{2025arXiv250918302P}
{Prieto-Lyon}, G., {Mason}, C.~A., {Strait}, V., {et~al.} 2025, arXiv e-prints,
  arXiv:2509.18302

\bibitem[{{Rhoads} {et~al.}(2000){Rhoads}, {Malhotra}, {Dey}, {Stern},
  {Spinrad}, \& {Jannuzi}}]{rhoads00}
{Rhoads}, J.~E., {Malhotra}, S., {Dey}, A., {et~al.} 2000, \apjl, 545, L85

\bibitem[{{Rhoads} {et~al.}(2004){Rhoads}, {Xu}, {Dawson}, {Dey}, {Malhotra},
  {Wang}, {Jannuzi}, {Spinrad}, \& {Stern}}]{rhoads04}
{Rhoads}, J.~E., {Xu}, C., {Dawson}, S., {et~al.} 2004, \apj, 611, 59

\bibitem[{{Rieke} {et~al.}(2023){Rieke}, {Robertson}, {Tacchella}, {Hainline},
  {Johnson}, {Hausen}, {Ji}, {Willmer}, {Eisenstein}, {Pusk{\'a}s}, {Alberts},
  {Arribas}, {Baker}, {Baum}, {Bhatawdekar}, {Bonaventura}, {Boyett}, {Bunker},
  {Cameron}, {Carniani}, {Charlot}, {Chevallard}, {Chen}, {Curti},
  {Curtis-Lake}, {Danhaive}, {DeCoursey}, {Dressler}, {Egami}, {Endsley},
  {Helton}, {Hviding}, {Kumari}, {Looser}, {Lyu}, {Maiolino}, {Maseda},
  {Nelson}, {Rieke}, {Rix}, {Sandles}, {Saxena}, {Sharpe}, {Shivaei},
  {Skarbinski}, {Smit}, {Stark}, {Stone}, {Suess}, {Sun}, {Topping},
  {{\"U}bler}, {Villanueva}, {Wallace}, {Williams}, {Willott}, {Whitler},
  {Witstok}, \& {Woodrum}}]{jades23}
{Rieke}, M.~J., {Robertson}, B., {Tacchella}, S., {et~al.} 2023, \apjs, 269, 16

\bibitem[{{Rinaldi} {et~al.}(2022){Rinaldi}, {Caputi}, {van Mierlo}, {Ashby},
  {Caminha}, \& {Iani}}]{rinaldi22}
{Rinaldi}, P., {Caputi}, K.~I., {van Mierlo}, S.~E., {et~al.} 2022, \apj, 930,
  128

\bibitem[{{Rivera-Thorsen} {et~al.}(2015){Rivera-Thorsen}, {Hayes},
  {{\"O}stlin}, {Duval}, {Orlitov{\'a}}, {Verhamme}, {Mas-Hesse}, {Schaerer},
  {Cannon}, {Ot{\'\i}-Floranes}, {Sandberg}, {Guaita}, {Adamo}, {Atek},
  {Herenz}, {Kunth}, {Laursen}, \& {Melinder}}]{riveraThorsen15}
{Rivera-Thorsen}, T.~E., {Hayes}, M., {{\"O}stlin}, G., {et~al.} 2015, \apj,
  805, 14

\bibitem[{{Roberts-Borsani} {et~al.}(2021){Roberts-Borsani}, {Treu}, {Mason},
  {Schmidt}, {Jones}, \& {Fontana}}]{robertsBorsani21}
{Roberts-Borsani}, G., {Treu}, T., {Mason}, C., {et~al.} 2021, \apj, 910, 86

\bibitem[{{Runnholm} {et~al.}(2025){Runnholm}, {Hayes}, {Mehta}, {Malkan},
  {Scarlata}, {Nedkova}, {Rafelski}, {Vulcani}, {Huberty}, {Herenz}, {Hutter},
  {Bruton}, {Acharyya}, {Atek}, {Baronchelli}, {Battisti}, {Brada{\v{c}}},
  {Bunker}, {Dai}, {Hannahs}, {Hasan}, {Kim}, {Leethochawalit}, {Lin},
  {Rutkowski}, {Saldana-Lopez}, {Sattari}, \& {Wang}}]{runnholm25}
{Runnholm}, A., {Hayes}, M.~J., {Mehta}, V., {et~al.} 2025, \apj, 984, 95

\bibitem[{{Saldana-Lopez} {et~al.}(2023){Saldana-Lopez}, {Schaerer},
  {Chisholm}, {Calabr{\`o}}, {Pentericci}, {Cullen}, {Saxena}, {Amor{\'\i}n},
  {Carnall}, {Fontanot}, {Fynbo}, {Guaita}, {Hathi}, {Hibon}, {Ji}, {McLeod},
  {Pompei}, \& {Zamorani}}]{saldanaLopez23}
{Saldana-Lopez}, A., {Schaerer}, D., {Chisholm}, J., {et~al.} 2023, \mnras,
  522, 6295

\bibitem[{{Sarrouh} {et~al.}(2026){Sarrouh}, {Asada}, {Martis}, {Willott},
  {Iyer}, {Noirot}, {Muzzin}, {Sawicki}, {Brammer}, {Desprez},
  {Rihtar{\v{s}}i{\v{c}}}, {Zabl}, {Abraham}, {Brada{\v{c}}}, {Doyon},
  {Antwi-Danso}, {Berek}, {Brown}, {Estrada-Carpenter}, {Favaro}, {Felicioni},
  {Forrest}, {Gaspar}, {Gould}, {Gledhill}, {Harshan}, {Jahan}, {Jagga},
  {Jude{\v{z}}}, {Marchesini}, {Markov}, {Matharu}, {MacFarland}, {Merchant},
  {M{\'e}rida}, {Mowla}, {Myers}, {Omori}, {Pacifici}, {Ravindranath},
  {Robbins}, {Strait}, {Sok}, {Tan}, {Tripodi}, {Wilson}, \&
  {Withers}}]{canucsTDR1}
{Sarrouh}, G. T.~E., {Asada}, Y., {Martis}, N.~S., {et~al.} 2026, \apjs, 282, 3

\bibitem[{{Shimizu} {et~al.}(2025){Shimizu}, {Kashikawa}, {Kikuta}, {Takeda},
  {Arita}, {Emori}, \& {Koretomo}}]{shimizu25}
{Shimizu}, S., {Kashikawa}, N., {Kikuta}, S., {et~al.} 2025, \mnras,
  arXiv:2506.03242

\bibitem[{{Simmonds} {et~al.}(2023){Simmonds}, {Tacchella}, {Maseda},
  {Williams}, {Baker}, {Witten}, {Johnson}, {Robertson}, {Saxena}, {Sun},
  {Witstok}, {Bhatawdekar}, {Boyett}, {Bunker}, {Charlot}, {Curtis-Lake},
  {Egami}, {Eisenstein}, {Ji}, {Maiolino}, {Sandles}, {Smit}, {{\"U}bler}, \&
  {Willott}}]{simmonds23}
{Simmonds}, C., {Tacchella}, S., {Maseda}, M., {et~al.} 2023, \mnras, 523, 5468

\bibitem[{{Sobral} \& {Matthee}(2019)}]{sobral19a}
{Sobral}, D., \& {Matthee}, J. 2019, \aap, 623, A157

\bibitem[{{Spina} {et~al.}(2024){Spina}, {Bosman}, {Davies}, {Gaikwad}, \&
  {Zhu}}]{spina24}
{Spina}, B., {Bosman}, S. E.~I., {Davies}, F.~B., {Gaikwad}, P., \& {Zhu}, Y.
  2024, \aap, 688, L26

\bibitem[{{Steidel} {et~al.}(1996){Steidel}, {Giavalisco}, {Pettini},
  {Dickinson}, \& {Adelberger}}]{steidel96a}
{Steidel}, C.~C., {Giavalisco}, M., {Pettini}, M., {Dickinson}, M., \&
  {Adelberger}, K.~L. 1996, \apjl, 462, L17

\bibitem[{{Tang} {et~al.}(2024{\natexlab{a}}){Tang}, {Stark}, {Topping},
  {Mason}, \& {Ellis}}]{tangm24b}
{Tang}, M., {Stark}, D.~P., {Topping}, M.~W., {Mason}, C., \& {Ellis}, R.~S.
  2024{\natexlab{a}}, \apj, 975, 208

\bibitem[{{Tang} {et~al.}(2024{\natexlab{b}}){Tang}, {Stark}, {Ellis}, {Sun},
  {Topping}, {Robertson}, {Tacchella}, {Arribas}, {Baker}, {Bhatawdekar},
  {Boyett}, {Bunker}, {Charlot}, {Chen}, {Chevallard}, {Jones}, {Kumari},
  {Lyu}, {Maiolino}, {Maseda}, {Saxena}, {Whitler}, {Williams}, {Willott}, \&
  {Witstok}}]{tangm24a}
{Tang}, M., {Stark}, D.~P., {Ellis}, R.~S., {et~al.} 2024{\natexlab{b}},
  \mnras, 531, 2701

\bibitem[{{Taniguchi} {et~al.}(2005){Taniguchi}, {Ajiki}, {Nagao}, {Shioya},
  {Murayama}, {Kashikawa}, {Kodaira}, {Kaifu}, {Ando}, {Karoji}, {Akiyama},
  {Aoki}, {Doi}, {Fujita}, {Furusawa}, {Hayashino}, {Iwamuro}, {Iye},
  {Kobayashi}, {Kodama}, {Komiyama}, {Matsuda}, {Miyazaki}, {Mizumoto},
  {Morokuma}, {Motohara}, {Nariai}, {Ohta}, {Ohyama}, {Okamura}, {Ouchi},
  {Sasaki}, {Sato}, {Sekiguchi}, {Shimasaku}, {Tamura}, {Umemura}, {Yamada},
  {Yasuda}, \& {Yoshida}}]{taniguchi05}
{Taniguchi}, Y., {Ajiki}, M., {Nagao}, T., {et~al.} 2005, \pasj, 57, 165

\bibitem[{{Tilvi} {et~al.}(2010){Tilvi}, {Rhoads}, {Hibon}, {Malhotra}, {Wang},
  {Veilleux}, {Swaters}, {Probst}, {Krug}, {Finkelstein}, \&
  {Dickinson}}]{tilvi10}
{Tilvi}, V., {Rhoads}, J.~E., {Hibon}, P., {et~al.} 2010, \apj, 721, 1853

\bibitem[{{Topping} {et~al.}(2024){Topping}, {Stark}, {Senchyna}, {Plat},
  {Zitrin}, {Endsley}, {Charlot}, {Furtak}, {Maseda}, {Smit}, {Mainali},
  {Chevallard}, {Molyneux}, \& {Rigby}}]{topping24}
{Topping}, M.~W., {Stark}, D.~P., {Senchyna}, P., {et~al.} 2024, \mnras, 529,
  3301

\bibitem[{{Tripodi} {et~al.}(2026){Tripodi}, {Napolitano}, {Pentericci},
  {P{\'e}rez-D{\'\i}az}, {Bhagwat}, {D'Eugenio}, {Arevalo-Gonzalez},
  {Arroyo-Polonio}, {Calabr{\`o}}, {Ciardi}, {Dickinson}, {Ferguson},
  {Gandolfi}, {Hirschmann}, {Hu}, {Koekemoer}, {Llerena}, {Lucas}, {Oey},
  {Papovich}, {Yung}, \& {Wang}}]{2026arXiv260306409T}
{Tripodi}, R., {Napolitano}, L., {Pentericci}, L., {et~al.} 2026, arXiv
  e-prints, arXiv:2603.06409

\bibitem[{{Valentino} {et~al.}(2023){Valentino}, {Brammer}, {Gould}, {Kokorev},
  {Fujimoto}, {Jespersen}, {Vijayan}, {Weaver}, {Ito}, {Tanaka}, {Ilbert},
  {Magdis}, {Whitaker}, {Faisst}, {Gallazzi}, {Gillman}, {Gim{\'e}nez-Arteaga},
  {G{\'o}mez-Guijarro}, {Kubo}, {Heintz}, {Hirschmann}, {Oesch}, {Onodera},
  {Rizzo}, {Lee}, {Strait}, \& {Toft}}]{valentino23}
{Valentino}, F., {Brammer}, G., {Gould}, K. M.~L., {et~al.} 2023, \apj, 947, 20

\bibitem[{{{\v{D}}urov{\v{c}}{\'\i}kov{\'a}}
  {et~al.}(2024){{\v{D}}urov{\v{c}}{\'\i}kov{\'a}}, {Eilers}, {Chen},
  {Satyavolu}, {Kulkarni}, {Simcoe}, {Keating}, {Haehnelt}, \&
  {Ba{\~n}ados}}]{durovcikova24}
{{\v{D}}urov{\v{c}}{\'\i}kov{\'a}}, D., {Eilers}, A.-C., {Chen}, H., {et~al.}
  2024, \apj, 969, 162

\bibitem[{{Vikaeus} {et~al.}(2024){Vikaeus}, {Zackrisson}, {Wilkins},
  {Nabizadeh}, {Kokorev}, {Abdurro'uf}, {Bradley}, {Coe}, {Dayal}, \&
  {Ricotti}}]{vikaeus24}
{Vikaeus}, A., {Zackrisson}, E., {Wilkins}, S., {et~al.} 2024, \mnras, 529,
  1299

\bibitem[{{Wang} {et~al.}(2023){Wang}, {Fujimoto}, {Labb{\'e}}, {Furtak},
  {Miller}, {Setton}, {Zitrin}, {Atek}, {Bezanson}, {Brammer}, {Leja}, {Oesch},
  {Price}, {Chemerynska}, {Cutler}, {Dayal}, {van Dokkum}, {Goulding},
  {Greene}, {Fudamoto}, {Khullar}, {Kokorev}, {Marchesini}, {Pan}, {Weaver},
  {Whitaker}, \& {Williams}}]{wangbj23}
{Wang}, B., {Fujimoto}, S., {Labb{\'e}}, I., {et~al.} 2023, \apjl, 957, L34

\bibitem[{{Wang} {et~al.}(2026){Wang}, {Wang}, {Zhou}, {Yang}, {Ji}, {Pang},
  {Tsai}, {Inoue}, {Tang}, {Nanayakkara}, {Glazebrook}, {Zhan}, \&
  {Chen}}]{2026arXiv260305907W}
{Wang}, S., {Wang}, X., {Zhou}, H., {et~al.} 2026, arXiv e-prints,
  arXiv:2603.05907

\bibitem[{{Wang} {et~al.}(2024){Wang}, {Wylezalek}, {De Breuck}, {Vernet},
  {Rupke}, {Zakamska}, {Vayner}, {Lehnert}, {Nesvadba}, \& {Stern}}]{wangwj24}
{Wang}, W., {Wylezalek}, D., {De Breuck}, C., {et~al.} 2024, \aap, 683, A169

\bibitem[{{Willott} {et~al.}(2022){Willott}, {Doyon}, {Albert}, {Brammer},
  {Dixon}, {Muzic}, {Ravindranath}, {Scholz}, {Abraham}, {Artigau},
  {Brada{\v{c}}}, {Goudfrooij}, {Hutchings}, {Iyer}, {Jayawardhana}, {LaMassa},
  {Martis}, {Meyer}, {Morishita}, {Mowla}, {Muzzin}, {Noirot}, {Pacifici},
  {Rowlands}, {Sarrouh}, {Sawicki}, {Taylor}, {Volk}, \& {Zabl}}]{willott22}
{Willott}, C.~J., {Doyon}, R., {Albert}, L., {et~al.} 2022, \pasp, 134, 025002

\bibitem[{{Witstok} {et~al.}(2025){Witstok}, {Maiolino}, {Smit}, {Jones},
  {Bunker}, {Helton}, {Johnson}, {Tacchella}, {Saxena}, {Arribas},
  {Bhatawdekar}, {Boyett}, {Cameron}, {Cargile}, {Carniani}, {Charlot},
  {Chevallard}, {Curti}, {Curtis-Lake}, {D'Eugenio}, {Eisenstein}, {Hainline},
  {Hausen}, {Kumari}, {Laseter}, {Maseda}, {Rieke}, {Robertson}, {Scholtz},
  {Shivaei}, {Williams}, {Willmer}, \& {Willott}}]{witstok25}
{Witstok}, J., {Maiolino}, R., {Smit}, R., {et~al.} 2025, \mnras, 536, 27

\bibitem[{{Yang} {et~al.}(2017){Yang}, {Malhotra}, {Gronke}, {Rhoads},
  {Leitherer}, {Wofford}, {Jiang}, {Dijkstra}, {Tilvi}, \& {Wang}}]{yang17b}
{Yang}, H., {Malhotra}, S., {Gronke}, M., {et~al.} 2017, \apj, 844, 171

\bibitem[{{Zhao} {et~al.}(2024){Zhao}, {Huang}, {He}, {Montero-Camacho}, {Liu},
  {Renard}, {Tang}, {Verdier}, {Xu}, {Yang}, {Yu}, {Zhang}, {Zhao}, {Zhou},
  {He}, {Kneib}, {Li}, {Li}, {Wang}, {Xianyu}, {Zhang}, {Gsponer}, {Li},
  {Rocher}, {Zou}, {Tan}, {Huang}, {Wang}, {Li}, {Rombach}, {Dong},
  {Forero-Sanchez}, {Ning}, {Shan}, {Wang}, {Li}, {Zhai}, {Wang}, {Zhao},
  {Shi}, {Mao}, {Huang}, {Guo}, \& {Cai}}]{2024arXiv241107970Z}
{Zhao}, C., {Huang}, S., {He}, M., {et~al.} 2024, arXiv e-prints,
  arXiv:2411.07970

\bibitem[{{Zhu} {et~al.}(2026){Zhu}, {Zheng}, {Bian}, {Yuan}, {Jiang}, {Zhang},
  {Lin}, \& {Guo}}]{2026arXiv260301487Z}
{Zhu}, S., {Zheng}, Z.-Y., {Bian}, F., {et~al.} 2026, arXiv e-prints,
  arXiv:2603.01487

\bibitem[{{Zhu} {et~al.}(2024){Zhu}, {Becker}, {Bosman}, {Cain}, {Keating},
  {Nasir}, {D'Odorico}, {Ba{\~n}ados}, {Bian}, {Bischetti}, {Bolton}, {Chen},
  {D'Aloisio}, {Davies}, {Davies}, {Eilers}, {Fan}, {Gaikwad}, {Greig},
  {Haehnelt}, {Kulkarni}, {Lai}, {Puchwein}, {Qin}, {Ryan-Weber}, {Satyavolu},
  {Spina}, {Walter}, {Wang}, {Wolfson}, \& {Yang}}]{zhuyd24}
{Zhu}, Y., {Becker}, G.~D., {Bosman}, S. E.~I., {et~al.} 2024, \mnras, 533, L49

\bibitem[{{Zibetti} {et~al.}(2009){Zibetti}, {Charlot}, \& {Rix}}]{zibetti09}
{Zibetti}, S., {Charlot}, S., \& {Rix}, H.-W. 2009, \mnras, 400, 1181

\end{thebibliography}
\end{document}